\documentclass{aa}

\usepackage{txfonts}
\usepackage{graphicx}
\usepackage{lscape}
\usepackage{longtable}

\begin{document}

\title{Monitoring luminous yellow massive stars in M33: \\ new yellow hypergiant candidates\thanks{Based  on  observations  made  with  the  Gran  Telescopio  Canarias (GTC), installed  in  the  Spanish  Observatorio  de  El  Roque  de  Los Muchachos of the Instituto de Astrofísica de Canarias, on the island of La Palma, and on data collected at Subaru Telescope, which is operated by the National Astronomical Observatory of Japan.}}
\author{M. Kourniotis\inst{1,2} \and A.Z. Bonanos\inst{1} \and W. Yuan\inst{3} \and L.M. Macri\inst{3} \and D. Garcia-Alvarez\inst{4,5,6} \and C.-H. Lee\inst{7}}

\institute{IAASARS, National Observatory of Athens, GR-15236 Penteli, Greece \\ \email{mkourniotis@astro.noa.gr,bonanos@astro.noa.gr} \and Section of Astrophysics, Astronomy and Mechanics, Faculty of Physics, National and Kapodistrian University of Athens, Panepistimiopolis, GR15784 Zografos, Athens, Greece \and Mitchell Institute for Fundamental Physics \& Astronomy, Department of Physics \& Astronomy, Texas A\&M University, College Station, TX 77843, USA \and Dpto.  de Astrof\'{i}sica, Universidad de La Laguna, 38206 La Laguna, Tenerife, Spain \and Grantecan CALP, 38712 Bre\~{n}a Baja, La Palma, Spain \and Instituto de Astrof\'{ı}sica de Canarias, E-38205 La Laguna, Tenerife, Spain \and Subaru Telescope, National Astronomical Observatory of Japan, 650 North Aohoku Place, Hilo, HI 96720, USA} 

\date{}
\authorrunning{Kourniotis et al.}
\titlerunning{YHGs in M33}

\abstract
{The evolution of massive stars surviving the red supergiant (RSG) stage remains unexplored due to the rarity of such objects. The yellow hypergiants (YHGs) appear to be the warm counterparts of post-RSG classes located near the Humphreys-Davidson upper luminosity limit, which are characterized by atmospheric instability and high mass-loss rates.}{We aim to increase the number of YHGs in M33 and thus to contribute to a better understanding of the pre-supernova evolution of massive stars.}{Optical spectroscopy of five dust-enshrouded yellow supergiants (YSGs) selected from mid-IR criteria was obtained with the goal of detecting evidence of extensive atmospheres. We also analyzed $BVI_{c}$ photometry for 21 of the most luminous YSGs in M33 spanning approximately nine years to identify changes in the spectral type that are expected based on the few well-studied YHGs. To explore the properties of circumstellar dust, we performed spectral energy distribution fitting of multi-band photometry of the 21 YSGs. We additionally conducted $K-$band spectroscopy of the YHG candidate B324 in search of processed ejected material.}{We find three luminous YSGs in our sample, stars 2, 6 and 13, with log~$L/L_{\odot} \gtrsim 5.35$ to be YHG candidates, as they are surrounded by hot dust and are enshrouded within extended, cold dusty envelopes. Our spectroscopy of star 2 shows emission of more than one H$\alpha$ component, as well as emission of \ion{Ca}{II} and [\ion{N}{II}], implying an extended atmospheric structure. In addition, the long-term monitoring of the star reveals a dimming in the visual light curve of amplitude larger than 0.5 mag that caused an apparent drop in the temperature that exceeded 500 K. We suggest the observed variability to be analogous to that of the Galactic YHG $\rho$ Cas. We further support the post-RSG classification of N125093 and B324 instead of being LBVs in outburst. Five less luminous YSGs are suggested as post-RSG candidates showing evidence of hot or/and cool dust emission.}{We demonstrate that mid-IR photometry, combined with optical spectroscopy and time-series photometry, provide a robust method for identifying candidate YHGs. Future discovery of YHGs in Local Group galaxies is critical for the study of the late evolution of intermediate-mass massive stars.}


\maketitle 

\section{Introduction}
\label{intro}

The upper part of the Hertzsprung-Russell (HR) diagram is home to the most enigmatic types of massive stars, which serve as progenitors of various subclasses of core-collapse supernovae (SNe). From the hot region occupied by luminous blue variables (LBV) and Wolf-Rayet (WR) stars to the coolest limit where red supergiants (RSGs) reside, the physics of stellar evolution is being refined. In these types of massive stars, the influence of mass loss is critical in modulating the evolutionary path and the lifetime of the star. OB-type stars  on the main sequence possess steady, line-driven and therefore metallicity-sensitive winds that are responsible for mass-loss rates of $\le 10^{-5}$ M$_\odot$ yr$^{-1}$ \citep{Puls08}. During the giant super-Eddington eruptions observed in LBVs, however, high mass-loss rates ($\ge 10^{-3}$ M$_\odot$ yr$^{-1}$) can only be justified by continuum-driven winds caused by electron scattering \citep{Owo04, Smith06}. Atmospheric inhomogeneity (porosity) due to optically thick structures in the atmosphere of LBVs has been shown to provide stability against the strong radiation field. On the other hand, RSGs are found to display slow dust-driven winds enhanced by radial pulsations \citep{Geh71,Bow88,vanL05}. Depending on metallicity and rotation, RSGs of initial mass $20-40$ M$_\odot$ exhibit high mass-loss rates that shift them back to warm temperatures, owing to the inability of the stars to sustain a convective H-rich outer envelope \citep{Ekstrom12, Smith14, Mey15}. \cite{Yoon10} propose a pulsationally-driven super-wind as a mechanism to drive the high mass-loss rates of RSGs with $M_{init}>19$ M$_\odot$. Their conclusion may explain the subclass of Type IIn SNe embedded within recently ejected dusty shells.

Yellow hypergiants \citep[YHGs;][]{deJ98} are post-RSG stars that occupy a region in the HR diagram between $4\,000-7\,000$ K. Their luminosity is confined above log~L/L$_{\odot}\sim5.4$ and below the Humphreys-Davidson limit \citep{Humph94} at log~L/L$_{\odot}=5.8$. Their most distinct characteristic is that they undergo several blueward loops against the cool limit of the instability zone known as the ``Yellow Void'' (hereafter, YV) \citep{deJag97, Nieu00}. Given the difficulty to accurately define the true temperature of the YHGs, the exact location of the YV is yet under debate with various values reported in the literature: $6\,000-9\,000$ K \citep{Humph02}, $7\,000-10\,000$ K \citep{deJ98}, and even $8\,500-12\,000$ K \citep{Sto12}. A star in the YV is characterized by a negative density gradient and very low or negative gravity (g$_\text{eff}<0.3$ cm~s$^{-2}$) as a result of an extented atmosphere \citep{Nieu95, deJ98,deJ01}. \cite{Nieu12} stressed that the YV comprises two separate instability regions that are associated with the ionization of H and He. A typical characteristic of the dynamic instability is the decrease of the volume-averaged value of the first generalized adiabatic exponent $<\Gamma_{1}>=<(\text{dln}P/\text{dln}\rho)_{\text{ad}}>$ below 4/3 in the outer stellar envelope \citep{Stoth96}. By investigating the stability of cool supergiant atmospheres, \cite{Lob01} demonstrated that $<\Gamma_{1}>$ assumes values below unity over a large geometrical fraction of the atmosphere of low-gravity Kurucz models with $6\,500\le T_{\text{eff}} \le 7\,500$ K. Unfortunately, observational evidence of dynamically unstable hypergiants is poor owing to their rarity and short-lived state. On their late blueward evolution, YHGs are believed to rapidly cross the YV and conclude their life as low-luminosity LBVs and WR-stars \citep{Maed82,Stoth96,Humph16}. The observational gap of LBVs with log~L/L$_{\odot}=5.7\pm0.1$, however, renders the evolution of post-YHGs with the same luminosity uncertain. These stars were suggested to evolve independently of an optically thick outer atmosphere and appear directly as WR-stars upon its dissipation \citep{Smith04}. On the other hand, \cite{Groh13} assigned a YHG classification to the progenitors of Type IIL/b SNe with initial masses 17-19 M$_\odot$. Their theoretical framework is further supported by the statement that, the yellow progenitors reported in the literature could actually be YHGs instead of yellow supergiants (YSGs). We note, however, that stars following this evolutionary channel have log~L/L$_{\odot}\lesssim5.2$, which is quite below the luminosity value of the Galactic YHGs.

Estimating the true evolutionary state of yellow, intermediate-mass massive stars remains a rather puzzling task, given that they cross regions on the HR diagram where typically a redward evolution takes place. According to \cite{deJ98}, the classification of a YHG requires broad emission of one or more H$\alpha$ components and broad Fraunhofer lines that are collectively known as the ``Keenan-Smolinski'' criteria. These criteria aim to distinguish slow-rotating supergiants from hypergiants, the latter of which show large velocities of tens of km s$^{-1}$ due to large-scale atmospheric turbulence. Owing to the difficulty in obtaining high-resolution spectroscopy of faint extragalactic objects, the literature of YHGs has been confined to the Milky Way and the Magellanic Clouds. \cite{Humph13} used the term ``warm hypergiants'' to characterize very luminous A and F--type stars in M31 and M33 showing broad H$\alpha$ emission typical of high mass loss, and emission of \ion{Ca}{II} and [\ion{Ca}{II}] indicative of a gaseous expanding shell in a low density circumstellar environment. In addition, an infrared excess may provide evidence for surrounding dust formed due to current or past mass loss events \citep{Gordon16}. Such events imply that the star has either already evolved as an RSG or has undergone dynamic instability that led to enhanced expulsion of gas. An overabudance of Na and N is supportive of the former scenario because it is indicative of material processed in the interior of an evolved star via the hot bottom burning process \citep{Quint13}. On timescales of years or even decades, YHGs display significant color variability associated with the collapse of atmospheric layers and the generation of pulsationally-driven shockwaves that cause expansion and cooling of the stellar photosphere \citep{Lob03}. An alternative scenario suggests that, during outbursts, enhanced mass-loss outflows form optically thick winds (pseudo-photospheres) similar to those of LBVs in outburst, which in turn yield changes in the apparent spectral type of the star \citep{Sto12,Nieu12}. Although YHGs share their spectral properties with LBVs and B[e] supergiants, their late spectral types and the presence of warm circumstellar dust were recently suggested by \cite{Humph16b} as a way to differentiate YHGs from the other two classes. On the other hand, the Galactic YHG $\rho$ Cas shows little and gradually dissipating dust excess, which emerges from a shell that was ejected during the historic eruption of 1946 \citep{Gil70,Shen16}. Similarly, the spectroscopic twin of $\rho$ Cas, the hypergiant HR8752, also lacks infrared excess \citep{Gil70}. Continued photometric monitoring along with high-resolution spectroscopy is therefore suggested as essential for an accurate YHG classification.

Due to their rarity, there are approximately fifteen reported YHGs in the Galaxy and the Magellanic Clouds \citep{deJ98}. Six additional Galactic YHGs in the young, open cluster Westerlund 1 were reported by \cite{Clark05}. Although all known YHGs display typical spectroscopic characteristics of their class, only a few were found to display variability in spectral type when systematically observed over time. Of these, the Galactic IRC+10420 is associated with extended nebulosity \citep{Tiff10}. The circumstellar environment of the star exhibits knots, arcs, shells and jet-like structures, and was shaped during two epochs of increased mass-loss rate of $10^{-3}-10^{-4}$ M$_{\odot}$ yr$^{-1}$ in the last 6\,000 years \citep{Shen16}. IRC+10420 is believed to be crossing the YV \citep{Humph02} to evolve as a low-luminosity LBV \citep{Humph16}. It has changed its apparent spectral type from F8I+ to early A in 30 years with an increase of $\sim120$ K per year \citep{Oudm98,Klock02}. $\rho$ Cas, the most well-studied YHG with a current spectral type of F8$-$G2, aroused interest after its outburst in 2000 during which $T_{\text{eff}}$ decreased by $\sim3\,000$ K within 200 days. The outburst was followed by a drop in the $V-$band light curve by almost one order of magnitude and the appearance of titanium-oxide bands in the optical spectrum arising from a cool, expanding shell. The estimated mass loss of the eruptive event is comparable to the mass loss of the eruptions of LBVs \citep{Lob03}.

In M33, Var A \citep{Hubble53,Humph87,Humph06} is a post-RSG object known to have undergone a high mass-loss episode of the order 10$^{-4}$ M$_\odot$ yr$^{-1}$, that lasted for almost 50 years. The outburst caused a change in the apparent spectral type from F to M and a large infrared excess at 10 $\mu$m due to the formation of dust. More recently, B324 and N125093 were reported by \cite{Humph13} to exhibit spectral features of YHGs, meaning broad asymmetric H$\alpha$ wings, \ion{Ca}{II} triplet and [\ion{Ca}{II}] in emission. Owing to their high luminosity that places them above the empirical Humphreys-Davidson limit, these stars were classified as LBVs in eruption \citep{Clark12,Valeev10}. However, \cite{Humph13} argued that their actual evolutionary state is uncertain, given that they strongly resemble the bona fide YHG IRC+10420 and that their luminosity could be overestimated due to the uncertainty in the adopted distance. 

In the current study, we aim to increase the number of YHG candidates in M33. We obtained optical spectroscopy for five luminous YSGs that show infrared excess, with the goal of detecting signatures of expanding shells, which are associated with a post-RSG evolutionary state. Additionally, we exploited three-band, long-term photometry of luminous, spectroscopically confirmed YSGs to search for variations in the spectral type. Near-infrared spectroscopy was obtained for two luminous objects to provide further evidence of their evolved nature. The paper is organized as follows: in Sect. \ref{sample} we focus on the selection of our sample and describe the photometric and spectroscopic observations, in Sect. \ref{spect} we discuss the five optical spectra, in Sect. \ref{temp} we describe the photometric monitoring and present results on the variability in temperature, in Sect. \ref{seds} we employ multi-band photometry to build and fit the spectral energy distributions (SEDs) of our studied stars, in Sect. \ref{disc} we discuss the evolutionary state of our most luminous targets and provide a detailed description of the new YHG candidate star 2, and in Sect. \ref{concl} we briefly summarize the results of the present work.

\section{Selected sample and observations}
\label{sample}

The most luminous YSGs occupy the same region on the HR diagram as their post-RSG counterparts, YHGs. We selected candidate YHGs among the sample of candidate YSGs in M33 reported by \cite{Drout12}, as objects with log~$L/L_{\odot}>5$. Our luminosity threshold is below the expected boundary for a YHG because the $L$ value can be underestimated due to uncertainty in the distance and the local extinction. In addition, log~$L/L_{\odot}\sim5$ corresponds approximately to an upper theoretical limit for the instability strip \citep{Ekstrom12} and hence prevents the selection of Cepheids. The work of \cite{Drout12} exploited the optical photometry from the Local Group Galaxies Survey \citep[LGGS;][]{Massey06} to detect YSGs in M33 based on their $B-V$ color. We selected 53 bona fide YSGs with log~L/L$_{\odot}>5$ that were flagged by \cite{Drout12} as ``rank 1'' stars showing velocities consistent to that of M33 and a strong \ion{O}{I} feature typical of supergiants. During the preparation of this manuscript, \cite{Massey16} and \cite{Gordon16} spectroscopically confirmed 17 and 29 YSGs in our sample, respectively. The latter study further suggested 12 YSGs from our list as post-RSG candidates, showing mass loss via stellar winds and/or surrounding dust. In total, 36 of the 53 YSGs were assigned an AFG spectral type from the literature.

We matched our sample to the 2MASS catalog \citep{Cutri03} and all but two sources yielded near-infrared (NIR) counterparts.  To provide evidence of circumstellar dust, our objects were matched to the catalog of $\sim50\,000$ mid-infrared (MIR) point sources in M33 provided by \cite{Thomp09}, obtained from multi-epoch imaging with \textit{Spitzer}. Adopting a search radius of 1$\arcsec$, we found 40 stars detected at both 3.6 $\mu$m and 4.5 $\mu$m. In addition, we matched the 53 YSGs with the WISE catalog \citep{Wri10} using a 6$\arcsec$ radius equal to the precision of WISE at 3.4, 4.6, and 12 $\mu$m, and noted whether two or more 2MASS or \textit{Spitzer} sources were detected within 6$\arcsec$. Finally, the targets were checked against variability surveys in M33 by \cite{Shpo06} in the visual and \cite{Quin07} in the infrared. A search against variable sources from the catalog of \cite{Hart06} yielded no matches. The resulting near- and mid-infrared photometry with uncertainties and reported spectral types of the 53 stars is provided in Table \ref{tab01}. We indicate cases where photometry corresponds to an upper limit with a `U'. The stars are sorted by decreasing luminosity from \cite{Drout12} and for brevity, designation based on this order is hereafter used. 

\subsection{Photometric observations}
\label{phsample}

The central part of M33 was monitored in $B,V,I_{c}$ as part of the DIRECT project \citep[and references therein]{Stan98,Bon03}, which constituted the first large-area CCD-based synoptic survey of M33. The corresponding database provided by \cite{Macri01} was matched with our list of luminous YSGs and light curves were complemented with $BVI_{c}$ observations taken between 2002 and 2006, resulting in a time span of approximately nine years. A detailed description of the latter photometry is presented in \cite{Pelle11}. Of the 53 YSGs, we excluded those stars with a $V-$band J variability index \citep{Stets96} lower than 0.75, and eventually studied 21 stars with available light curves in all three filters. The resulting photometry on average consists of 40, 150, 90 datapoints in the $B-,V-$ and $I_{c}-$band, respectively. The time-series photometry of the 21 YSGs is available in its entirety in a machine-readable form at the CDS. A portion that contains the first three $B-$band measurements of star 1 is shown in Table~\ref{tab02} for guidance regarding its form and content.

\subsection{Spectroscopic observations}
\label{spsample}

\begin{figure*} 
\centering
\includegraphics[width=7.5in]{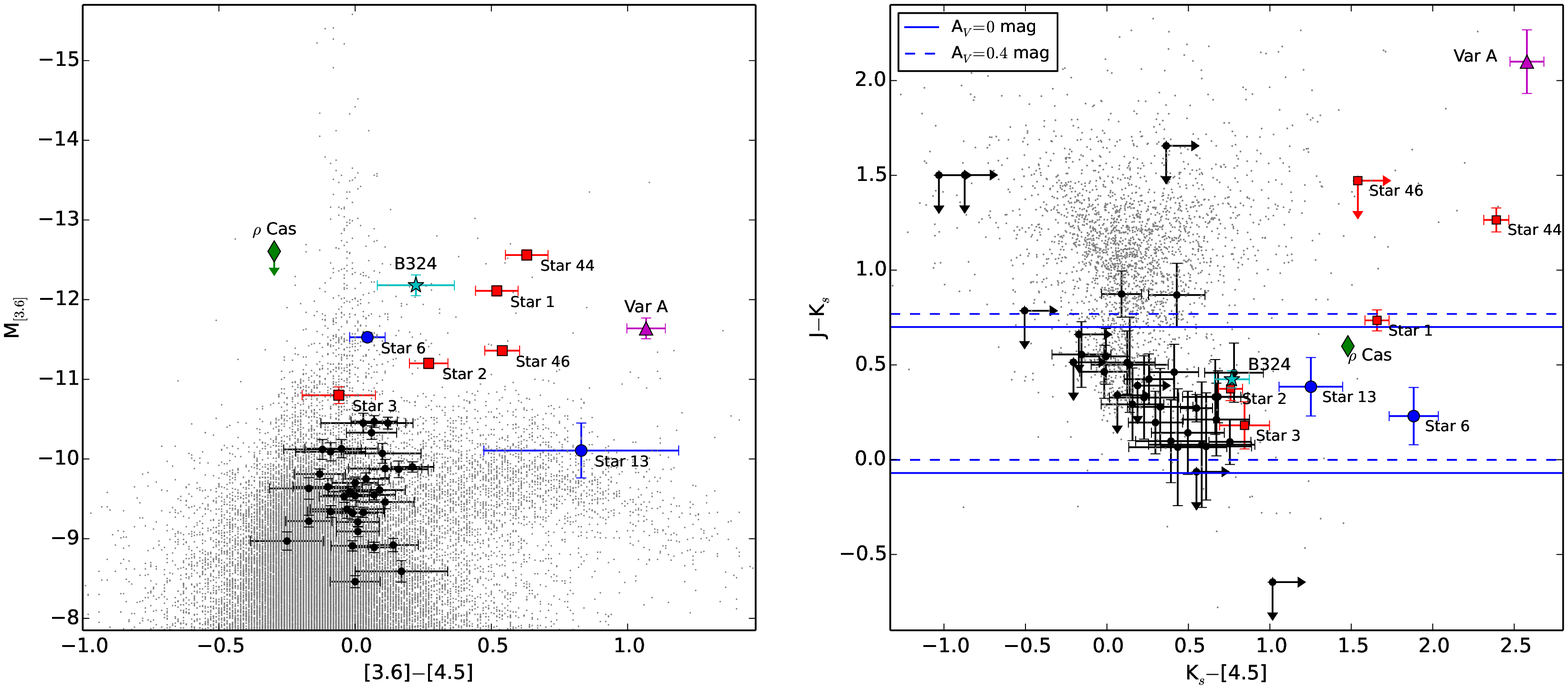}
\caption{Color$-$magnitude \textit{(left)} and color$-$color \textit{(right)} diagrams for the sample of 53 luminous YSGs in M33. Infrared photometry is taken from 2MASS and from \cite{Thomp09}. Optical spectroscopy with GTC was obtained for the five dust-enshrouded YSGs shown with red squares, following the designation in Table \ref{tab01}. Stars 6 and 13, which are flagged as candidate YHGs in the current work, are also plotted (large blue circles) using photometry from WISE at 3.4 and 4.6 $\mu$m. The $J-K_{s}$ boundaries for YSGs are indicated by the two solid lines while the dashed lines correspond to the boundaries corrected for the average visual extinction $A_{V}=0.4$ mag of M33. We show the location on the diagrams of the known YHGs $\rho$ Cas (green diamond) in the Galaxy and Var A (purple triangle) in M33, as well as of the candidate YHG/LBV B\,324 (cyan star) in M33. Photometry for $\rho$ Cas is taken from WISE at 3.4 and 4.6 $\mu$m, corresponding to upper limits due to the brightness of the object. Conversion to M$_{[3.6]}$ is done using distance moduli of 24.92 mag for M33 \citep{Bonanos06} and $12.5$ mag for $\rho$ Cas \citep{Zso91}.}
\label{fig01}
\end{figure*}

We undertook spectroscopy of four dust-enshrouded targets, stars 1, 2, 44 and 46 (namely \object{LGGS J013349.86+303246.1}, \object{LGGS J013358.05+304539.9}, \object{LGGS J013415.42+302816.4}, \object{LGGS J013442.14+303216.0}) chosen as YSGs with MIR excess attributed to the presence of warm circumstellar dust. According to the roadmaps of \cite{Bon09,Bon10} for massive stars with known spectral types in the Large and Small Magellanic Cloud, YSGs typically display $[3.6]-[4.5]\le0.1$ mag. We plotted our sample on a MIR color-magnitude diagram (CMD) as shown in the left panel of Fig. \ref{fig01} and selected the four sources with $[3.6]-[4.5]>0.25$ mag, which, interestingly, are the most luminous YSGs at 3.6 $\mu$m with $M_{3.6}\le-11$ mag. Conversion to M$_{3.6}$ is done using a distance modulus of 24.92 mag for M33, derived from the study of an O$-$type detached eclipsing binary \citep{Bonanos06}. Furthermore, the $J-K_{s}$ color was plotted against the $K_{s}-[4.5]$ on a color-color (CCD) diagram shown in the right panel of Fig. \ref{fig01}, where our four selected sources display $K_{s}-[4.5]\ge0.76$ mag. Stars 6 and 13, which are reported as YHG candidates in the present study are also shown, using their available photometry from WISE at 3.4 and 4.6 $\mu$m. The whole sample resides within the $J-K_{s}$ values reported in \cite{Neu12} for YSGs, when corrected for the extinction of M33 \citep[$A_{V}=0.4$ mag;][]{Massey07} and using the relation $E(J-K)=0.535 \times E(B-V)$ \citep{Schle98}. We added a fifth spectroscopic target, star 3 (\object{LGGS J013411.32+304631.4}), with $K_{s}-[4.5]=0.85$ mag and $M_{3.6}\le-10.8$ mag. For comparison, in both the CMD and CCD of Fig. \ref{fig01} we show the location of the YHG Var A and the candidate YHG B324 in M33, as well as that of the Galactic YHG $\rho$ Cas, for which uncertainty in the photometry is not available due to its high brightness. For the latter object, a distance modulus of $12.5$ mag \citep{Zso91} is used to obtain the absolute magnitude.

\begin{figure*} 
\centering
\includegraphics[width=4.5in]{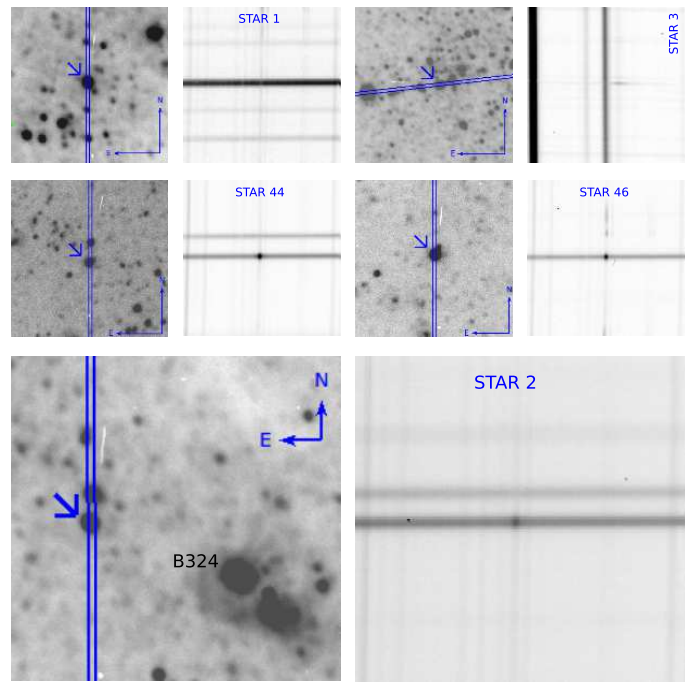}
\caption{Pairs of panels for the five targets observed with OSIRIS/GTC. In the left panel of each set, we show an 1$\arcmin$ x 1$\arcmin$ spatial view of the target in Sloan r$^\prime$ band. North is up, east is left. The slit in each case is shown with solid lines and targets are indicated with blue arrows. In the right panel of each set, we show the respective two-dimensional portion of the GTC spectrum of the object, centered on the H$\alpha$ line. We particularly emphasize star 2 (large set of panels) and indicate the candidate LBV/YHG B324, located $\sim30\arcsec$ away.}
\label{fig02}
\end{figure*}

\begin{figure*} 
\centering
\includegraphics[width=7in]{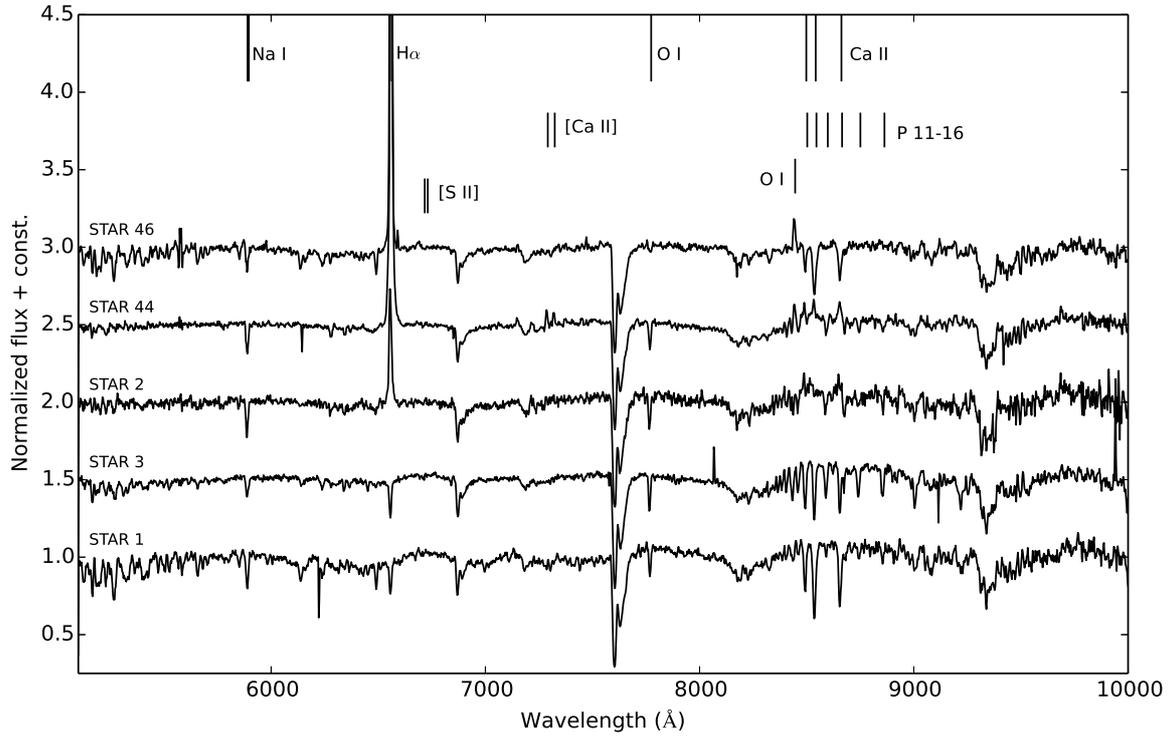}
\caption{Optical/infrared spectra obtained with OSIRIS/GTC of our five selected YSGs. For clarity, a constant of 0.5 is added to the normalized flux. The rest wavelengths of fundamental features are indicated on the top of the figure.}
\label{fig03}
\end{figure*}

Optical long-slit spectroscopy for the five dusty YSGs was obtained on UT 19 Aug 2014 with the Optical System for Imaging and low-intermediate-Resolution Integrated Spectroscopy (OSIRIS), installed on the Nasmyth-B focus of the segmented 10.4m Gran Telescopio Canarias (GTC). The instrument is equipped with a mosaic of two 2048 x 4102 detectors, with a 7.8$\arcmin$ x 7.8$\arcmin$ unvignetted field of view and a pixel scale of $0.254\arcsec$ pixel$^{-1}$ for the standard operation mode using a 2 x 2 binning. We selected the R1000R grism that covers the wavelength range $5\,100-10\,000$ $\AA$ and a slit width of $1.2\arcsec$ yielding a resolution R of approximately $700-750$ as measured from the full-width at half maximum (FWHM) of sky lines in the region of the \ion{Ca}{II} triplet ($\sim8\,500$ \AA). The slit was set along the parallactic angle (north-south), with an exception for star 3 in order to include a bright source located $25\arcsec$ east of the target. A 1$\arcmin$ x 1$\arcmin$ spatial view of each object and the orientation of the slit is shown in the left panels of Fig. \ref{fig02}. We additionally provide a two-dimensional view of the GTC spectrum centered on the H$\alpha$ feature in the right panels of Fig. \ref{fig02}, to evaluate the possibility of contamination of the line by extended nebular emission. Depending on the brightness of the objects in the $V-$band ($16-17.3$ mag), the exposure time ranged from 100 to 600 sec resulting in a signal-to-noise ratio (SNR) of $60-120$. A log of the observations is provided in Table \ref{tab03}. The five spectra were bias-subtracted and flat-normalized using basic IRAF routines\footnote{IRAF is distributed by the National Optical Astronomy Observatories, which are operated by the Association of Universities for Research in Astronomy, Inc., under cooperative agreement with the National Science Foundation.}. For the wavelength calibration, exposures of HgAr, Xe and Ne lamps were taken following the observations. A low-order cubic spline was used to normalize the spectra, which are displayed in Fig. \ref{fig03}. We similarly extracted the spectra of additional stars that were found in the slit during our observations, which turned out to be absorption-line spectra of foreground dwarfs or spectra with sufficiently low SNR to be accurately classified. 

\begin{figure*} 
\centering
\includegraphics[width=5.5in]{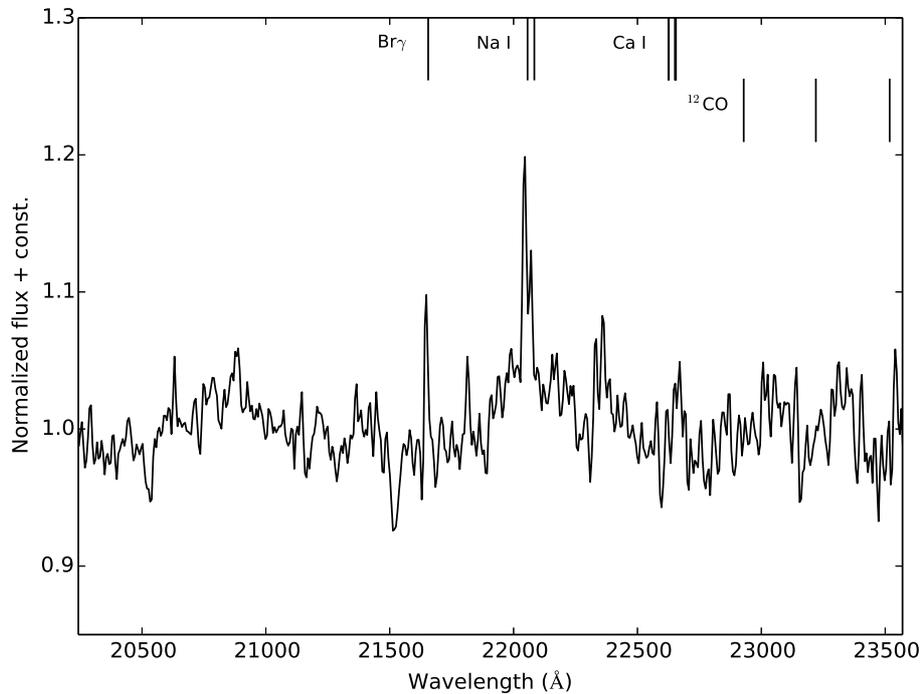}
\caption{Normalized near-infrared spectroscopy of the candidate YHG/LBV B\,324 observed with IRCS/Subaru. The rest wavelengths of discussed features are indicated on the top of the figure.}
\label{fig04}
\end{figure*}

We also undertook NIR spectroscopy of two objects in M33, namely star 2 and the candidate LBV/YHG B324. The former was among the five YSGs observed with GTC that turned out to be a rather promising target for understanding the late evolution of massive stars. Observations were taken on UT 5 Oct 2015, with the Infrared Camera and Spectrograph (IRCS) mounted on the Nasmyth focus of the 8.2m Subaru Telescope at Mauna Kea. The camera is equipped with a 1024 x 1024 detector and a $54\arcsec$ field of view. We used the $K-$band grism with a 52 mas pixel scale that covers the wavelength range $1.93-2.48$ $\mu$m. By measuring a FWHM of 13 $\AA$ for an unblended Argon feature at 21540 $\AA$ of the comparison lamp, we derive a resolution of $R\sim1700$. For an accurate sky subtraction, both objects were observed in an object-sky-object (OSO) sequence with a 300 second time per exposure. We conducted 1.5 OSO and six OSO sequences for the B324 and star 2, respectively. A log of the NIR observations is provided in Table \ref{tab03}. The spectra of each object were combined and wavelength-calibrated following the standard procedure with IRAF. To correct for the atmospheric transmittance, the bright late-B dwarf HD\,1606 was observed between the two objects and was similarly extracted as described above. The intrinsic Br$\gamma$ feature at 2.166 $\mu$m of the standard star was masked by interpolating the continuum level. We then used the task \textit{telluric} to optimally scale the standard spectrum and divide the spectra of the objects. Unfortunately, due to bad weather conditions we could not obtain a sufficiently long exposure for the faint star 2, which is the reason for the rather low SNR ($\sim15$) of the data. We therefore were not able to further analyze the spectrum of star 2. The $K-$band spectrum of B324 is shown in Fig. \ref{fig04}. We indicate the rest wavelengths for the spectral lines of Br$\gamma$, \ion{Na}{I}, \ion{Ca}{I}, and CO. Notably, emission of \ion{Na}{I} is visible in the spectrum of the object.

\section{Analysis of the optical spectra}
\label{spect}

\begin{figure*} 
\centering
\includegraphics[width=4.5in]{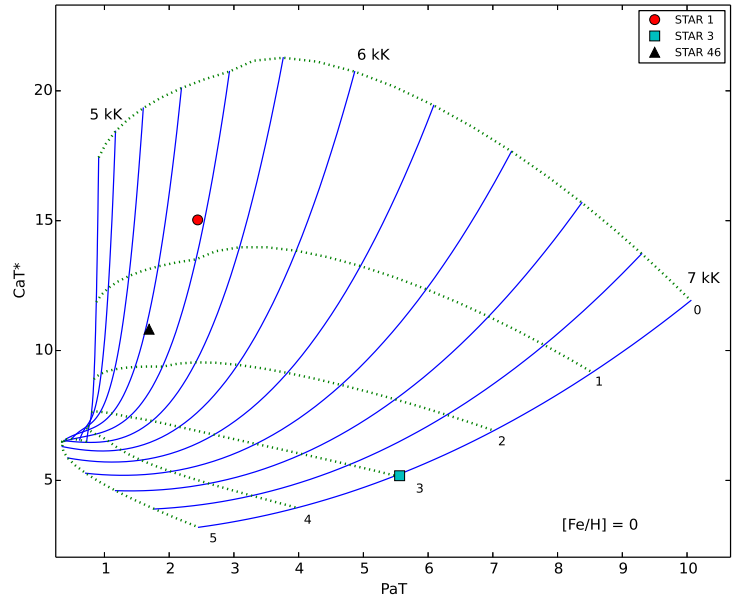}
\caption{Grid of atmospheric parameters as a function of the two generic indices CaT* and PaT, which measure the strength of \ion{Ca}{II} and Paschen lines, respectively. We show the positions on the grid of the three spectroscopic targets observed with the GTC, which show absorption in the \ion{Ca}{II} triplet. From left to right, solid lines indicate temperatures at the range $4\,800-7\,000$ K with a step of 200 K. Dotted lines correspond to log$g$ values of 0 to 5, with a step of 1 dex, as CaT* decreases.}
\label{fig05}
\end{figure*}

The observed optical wavelength range is essential for estimating the parameters of YSGs because it contains the \ion{Ca}{II} triplet at 8498.02, 8542.09 and 8662.14 $\AA$, which is prominent in stars of mid-F to early-M spectral type. For giants with temperatures above 5\,500 K, the red wings of the feature are significantly blended with hydrogen lines of the Paschen series. This temperature threshold decreases with increasing luminosity \citep{chmi00}. To determine the spectroscopic temperature for the three targets showing \ion{Ca}{II} in absorption, we employed the generic index CaT* \citep{Cen01} that measures the equivalent width of the integrated \ion{Ca}{II} triplet, corrected for the contamination from the adjacent Paschen lines. The strength of the latter is described by the index PaT and both indices for each star are provided in Table \ref{tab04}. We then estimated $T_{\text{eff}}$ and log$g$ by interpolating on a grid of stellar atmospheric parameters as a function of CaT* and PaT, shown in Fig. \ref{fig05}. These parameters were derived from the empirical fit using a NIR library of approximately 700 stars \citep{Cen02} at $\text{[Fe/H]}=0$. We note that the dependence on metallicity is stronger with increasing temperature. Employing models at $\text{[Fe/H]}=-0.5$, the inferred gravity and temperature for the two stars with $T_{\text{eff}}<6\,000$ K decrease by about 0.5 dex and 100 K, respectively, whereas for the hotter case, by almost 1 dex and 500~K, respectively. 

A close-up view of the \ion{Ca}{II} triplet in emission for stars 2 and 44 is shown in the left panel of Fig. \ref{fig06}. For comparison, we superpose two ATLAS9 synthetic spectra of 5\,500 K and 6\,250 K by \cite{Mun05}, at solar metallicity. The templates were broadened to match the resolution of our data, normalized and optimally blueshifted to fit the P14 hydrogen line at 8598.4 $\AA$. The temperature of the models is selected to match the SED-fit temperature discussed in Sect. \ref{seds}. We stress that, for stars exhibiting intense mass loss, hydrogen lines are not an optimal tracer of the stellar photosphere and are frequently filled in by emission, thus resulting in a pseudo-cooling of the effective temperature. Therefore, we do not attempt to estimate spectroscopic $T_{\text{eff}}$ values for these stars, because high-resolution spectroscopy of reliable photospheric lines (such as Fe) is needed.

The H$\alpha$ line in emission constitutes a fundamental mass-loss tracer, which allows inference on the wind structure of hot supergiants \citep{Petr04} and warm hypergiants \citep[e.g.][]{deJag97b}. In addition, the \ion{O}{I} feature at 7774 $\AA$ serves as an indicator for the luminosity of AFG type supergiants \citep{Arel03} and exhibits a strong dependence on the temperature for $T_{\text{eff}}<6\,000$ K \citep{Kovt12}. We derived radial velocities and equivalent widths for the \ion{O}{I} line and the central H$\alpha$ component by fitting gaussian functions with the IRAF \textit{splot} routine. The values are provided in Table \ref{tab04}. Errors were estimated using a Monte Carlo simulation with 100 samples, assuming a Gaussian sigma of the noise that was equal to the root-mean-square measured from continuum regions. The barycentric correction was calculated and the value of 25 km s$^{-1}$ was added to the measured velocities. We additionally provide values for $M_{V}$ based on the strength of the \ion{O}{I} 7774 $\AA$ blend, following the equations by \cite{Ferro03}. We infer luminosities using bolometric corrections from \cite{Flow96} for temperatures derived from the fit of the multi-band photometry, as discussed in Sect. \ref{seds}. We briefly describe each of the five spectra in the following paragraphs, whereas in Sect. \ref{disc} we discuss their evolutionary state.

\subsection*{Star 1}
The absorption spectrum of the star is typical of a yellow supergiant and both works by \cite{Massey16} and \cite{Gordon16} agree to a late-F or early-G spectral type. The strengths of the \ion{Ca}{II} and the Paschen series in the GTC spectrum indicate a $T_{\text{eff}}=5\,500-5\,600$ K and log$g\sim0.8$, as inferred from Fig. \ref{fig05}. We report disagreement between our radial velocity measurements and the value given in \cite{Drout12} by $70-100$ km~s$^{-1}$.

\subsection*{Star 2}
The obtained spectrum shows emission in the \ion{Ca}{II} triplet, which is expected in the spectra of YHGs, being indicative of gaseous circumstellar ejecta \citep{Humph13}. Emission is evident for H$\alpha$ and does not appear to extend north- and southwards of the target along the slit (Fig. \ref{fig02}), hence denoting a stellar origin due to mass loss. The absence of [\ion{S}{II}] at 6717 and 6731 $\AA$ in emission also renders nebular contamination less likely. We measure broad asymmetric wings at velocities of $-660\pm30$ and $520\pm30$ km s$^{-1}$ from the core of the line as shown in the right panel of Fig. \ref{fig06}. These values are unrealistic for wind outflow velocities, indicating a contribution from lines other than H$\alpha$. The blueshifted component is likely due to [\ion{N}{II}] emission. The [\ion{N}{II}]  contribution on the red wing contributes less flux, hinting toward the presence of a strong,
redshifted H$\alpha$ component. A high-resolution spectrum would clearly resolve the profile and allow measurement of the outflow velocity. We derive velocities for the central H$\alpha$ component and the \ion{O}{I} line of $-320\pm5$ and $-284\pm12$ km s$^{-1}$ respectively, which both deviate from the value of $-214$ km s$^{-1}$ measured by \cite{Drout12}. Nevertheless, our velocity measurement for \ion{O}{I} is in good agreement with the expected value of $-261$ km s$^{-1}$, which was provided by the latter study as a function of the position and the radial distance of the object from the center of M33. Discrepancies among the studies are likely to arise due to pulsational activity. The measured velocities of the individual \ion{Ca}{II} components yield blueshifted values of approximately $-420$ km s$^{-1}$ that are likely overestimated given that the emission is partly washed out by the absorption of the adjacent hydrogen lines. 

\subsection*{Star 3}
Interpolation of the inferred indices for the \ion{Ca}{II} triplet and the neighboring hydrogen lines from Fig. \ref{fig05} yields a $T_{\text{eff}}=7\,000$ K that agrees with the value of \cite{Drout12}. We find a large surface gravity of log$g$=3 that conflicts with the luminosity value log~$L/L_{\odot}>5.1$ measured from the strong \ion{O}{I} blend. The latter value of luminosity is also supported from the fit of the SED (Sect. \ref{seds}), which implies values of gravity log$g<$1.5. We attribute the discrepancy either to the low SNR of the spectrum or to the fact that the \ion{Ca}{II}/Paschen blend could be filled in by emission. We derive consistent radial velocities for the absorption of H$\alpha$ and \ion{O}{I} lines at $-275$ and $-280$ km s$^{-1}$, respectively, which do not conflict with the expected velocity of $-261$ km s$^{-1}$ for the position of the object in M33.

\subsection*{Star 44}
Known as N125093 in the literature, the star is suggested by \cite{Humph13} as a candidate YHG based on signatures in its spectrum of an expanding shell and a low density circumstellar environment. Moreover, emission due to the surrounding dust emerges in the infrared regime. We confirm the presence of \ion{Ca}{II} and that of [\ion{Ca}{II}] in our spectrum. We measure broad emission wings red- and blueward of the central H$\alpha$ component as displayed in Fig. \ref{fig06}, at velocities of $570\pm10$ and $-680\pm10$ km s$^{-1}$, respectively. As in the case of star 2, emission of [\ion{N}{II}] is blended with the wings of the H$\alpha$ line. The recent, moderate-resolution spectroscopy by \cite{Gordon16} confirms the presence of the [\ion{N}{II}] $\lambda6548,6583$ lines. The measured velocity of the main H$\alpha$ feature deviates by $\sim100$ km s$^{-1}$ with respect to that of \ion{O}{I}, indicating emission that arises in different regions of the star. 

\subsection*{Star 46}
The star displays emission of H$\alpha$ and a weak presence of blueshifted [\ion{S}{II}] at 6717 and 6731 $\AA$ arising from nebular contamination. We find evidence of H$\alpha$ emission $\sim7\arcsec$ north of the target as shown in Fig. \ref{fig02}. We do not report wings of the feature as in the two previous cases with hydrogen emission, although a weak redshifted emission feature could be present due to electron scattering. The weak absorption of \ion{O}{I} is due to the relatively low luminosity and temperature of the star, which is also confirmed from our fit of the multi-band photometry, as discussed in Sect. \ref{seds}. In addition, the instrinsic absorption of \ion{O}{I} could be masked due to the observed nebular emission of the \ion{O}{I} line at 8446 $\AA$. We do not provide a velocity measurement for the \ion{O}{I} 7774 $\AA$ line because it is affected by noise.

\section{Temperature variability}
\label{temp}

Having obtained the $BVI_{c}$ light curves of 21 luminous YSGs, we searched for possible variations in the spectral type of the objects on a timescale of approximately nine years. The data were binned into intervals of four days to discard possible outliers in the photometry but preserve short-term variations. For each bin, the mean magnitude was calculated, weighted by the uncertainty of the measurements. The standard deviation of the data points was adopted as a conservative value for the uncertainty of the binned measurement. We then proceeded to use only those bins that had both $B-V$, $V-I_{c}$ colors available and for each bin we determined $T_{\text{eff}}$ by fitting to synthetic models. 

The 21 YSGs with available time-series photometry are located within 17$\arcmin$ from the center of M33, corresponding to 4 kpc. At that radius, the global metallicity is similar to solar metallicity, based on the metallicity gradient derived by \cite{Urba05} with the use of B-type supergiants. We employed the updated ATLAS9 atmospheric models by \cite{How11} with a canonical turbulent-velocity parameter $v_{t}=2$ km s$^{-1}$, at solar metallicity. According to the stellar models by \cite{Ekstrom12}, our studied region of the upper HR diagram is occupied by models with log$g<1.5$, depending on the evolutionary state of the star. We used ATLAS9 models with log$g=0.5$ that range from 3500 K to 7500 K, while for stars showing higher temperatures we adopted models with log$g=1.0$ that reach 8750 K. For each $BVI_{c}$ bin, we ran a $\chi^{2}$-minimization algorithm over a Levenberg-Marquardt fitting scheme, leaving the temperature of the model and the scaling factor as free parameters. The color excess for each target was adopted based on colors of nearby early-type stars. In particular, we averaged the color excesses of all OB- and A-type stars from \cite{Massey16} within $0.5\arcmin$ from each target when available, and corrected for their intrinsic colors from \cite{Fitz70}. In the absence of such stars, we adopted the average reddening of M33, namely $E(B-V)=0.13\pm0.01$ mag \citep{Massey95}. Standard magnitudes were converted to fluxes using the zero points provided by \cite{Bess98}. The SEDs were reddened according to the extinction laws by \cite{Card89} and assuming a total-to-selective extinction ratio $R_{V}=3.2$. For each bin, we ran the fitting routine for 1000 different sets of the $BVI_{c}$ photometry and the reddening, which were randomly selected within their uncertainties. To account for the 250 K step of the Kurucz models, we ran a second bootstrap procedure on the derived sample of temperatures, randomly selecting them within $\pm125$ K. The resulting $T_{\text{eff}}$ values for every studied YSG are plotted in Fig. \ref{fig07}.

We compared our temperatures to the values provided by \cite{Drout12}, which were determined upon transformation of the $B-V$ color measured by \cite{Massey06} in Sep and Oct of 2000 and 2001, and we report discrepancies that occur mainly due to three reasons. First, \cite{Drout12} used lower-metallicity Kurucz models ($Z=0.6$ $Z_\odot$) that yield slightly cooler temperatures. Second, in our work we fit all $BVI-$bands whereas in \cite{Drout12} only $B-V$ was used to determine temperatures. Finally, intrinsic variability of the star may cause discrepancies.

When inspecting the calculated temperatures in Fig. \ref{fig07}, one may notice a dispersion of individual points that on average does not exceed $\sim250$ K from the baseline. This could be due to uncertainties in the photometry or even to low-amplitude variations arising from mechanisms such as pulsations \citep{Sto12}, the study of which is beyond the scope of this work. On the other hand, a systematic high-amplitude trend of the baseline could imply critical photospheric activity typical of the ``bounces'' of YHGs. We adopted the median absolute deviation (MAD) as a statistical method to define variability in our sample \citep{Soko17}. We measured the MAD of temperatures for each of the 21 lightcurves yielding a mean value of 40 K with a standard deviation 45 K. Stars 2 and 51 stand out with a significance above 2$\sigma$, with MAD values of 160 K and 140 K, respectively. Both stars display variations in the temperature with amplitude equal or larger than 500 K.

\section{Spectral energy distributions}
\label{seds}

We proceeded to investigate any possible connection between the parameters determined from the DIRECT monitoring and the properties of the surrounding dust indicated by the infrared excess. We built the SEDs of our 21 selected YSGs, complementing the 2MASS photometry with the infrared photometry from \textit{Spitzer} and WISE. To eliminate possible contamination from nearby sources in the infrared, we discarded the WISE photometry from the fit when more than two sources from 2MASS and/or \textit{Spitzer} were found within the matching radius of $6\arcsec$ and, at the same time, the coordinates of the WISE source deviated significantly ($\gtrsim5\arcsec$) from those of the optical source. Following the above criterion, the WISE photometry was not included for stars 14, 26, 29, and 40, for all of which it was also found to be in an evident disagreement with the similar IRAC  bands from \textit{Spitzer}. 

The flux $f_{\lambda}$ measured from a star with dust excess at a distance $D$, at a wavelength $\lambda$, reddened to extinction $A(\lambda)$, can be written as

\begin{equation}
$$ f_{\lambda}=\left(\dfrac{R}{D}\right)^2 \times F_{\lambda}(T_{\star}) \times 10^{-0.4A(\lambda)} + \sum_{j}^{n} C_{j}\lambda^{-\beta} B_{\lambda}(T_{j}) $$ 
\end{equation}

\noindent where $R$ is the stellar radius, $F_{\lambda}$ is the surface flux of the ATLAS9 model corresponding to an effective temperature $T_{\star}$, and $B_{\lambda}$ is the Planck function of a modified blackbody with an emissivity spectral index $\beta$, to approximately model the j-th dust component with temperature $T_{j}$ and scale factor $C_{j}$. We ran the Levenberg-Marquardt fitting-algorithm over the grid of models described in Sect. \ref{temp}, to fit the optical and infrared data of each target and obtain the radius and effective temperatures of the star and its dust components. The dust-emissivity index $\beta$ was set equal to the typical value of 1.5 and a distance of $964\pm54$ kpc \citep{Bonanos06} was adopted. For consistency, we averaged the DIRECT data of the observing window around JD 2\,450\,700, to obtain the optical $BVI_{c}$ counterpart that was closest to the date of the 2MASS observation ($\sim$ JD 2\,450\,790). The number $n$ of the dust components was chosen each time as the minimum to optimize the fit to the infrared data. The best-fit SEDs to the observed photometry are shown in Fig. \ref{fig08}. Bolometric luminosities were calculated by integrating the resulting composite SEDs, which were corrected for the interstellar extinction, up to 8 and 22 $\mu$m. Integration to 22 $\mu$m is separately provided to evaluate the contribution of possible MIR contamination from nearby PAH emission. As in Sect. \ref{temp}, uncertainties were derived using a bootstrap procedure that runs the fit routine for 1000 sets of the input parameters (photometry, reddening, and distance) randomly selected within their errors. We note that the luminosity error bars are dominated by the error in the adopted distance. The resulting effective temperatures, radii, bolometric luminosities, and dust temperatures, along with their 1$\sigma$-uncertainties, are provided in Table \ref{tab05}.

\begin{figure*} 
\centering
\includegraphics[width=6in]{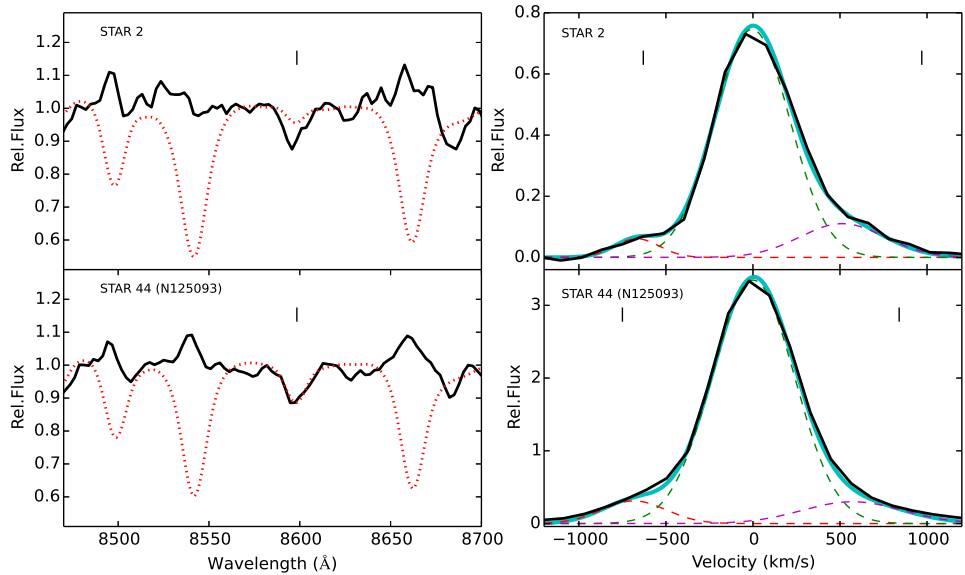}
\caption{\textit{Left.} Spectral region $8470-8700$ $\AA$ for stars 2 and 44, which show \ion{Ca}{II} in emission (thick black line). The vertical line marks the rest wavelength of the P14 hydrogen line from the Paschen series at 8598.4 $\AA$. Synthetic spectra from \cite{Mun05} corresponding to the SED-fit temperatures of 5\,500 and 6\,250 K are superposed on the spectra of stars 2 and 44, respectively, blueshifted to optimally fit the P14 line. \textit{Right.} Analysis of the H$\alpha$ feature of the same stars shown in the left panel. Red- and blue-shifted wings are indicated with dashed lines. The vertical lines indicate the velocities of the [\ion{N}{II}] $\lambda6548,6583$ lines relevant to the central H$\alpha$ component and corrected for the value inferred from the measurement of \ion{O}{I}. The total composite emission is shown with a thick blue line, fitting observations indicated with a black line.}
\label{fig06}
\end{figure*}

\begin{figure*} 
\centering
\includegraphics[width=6.3in]{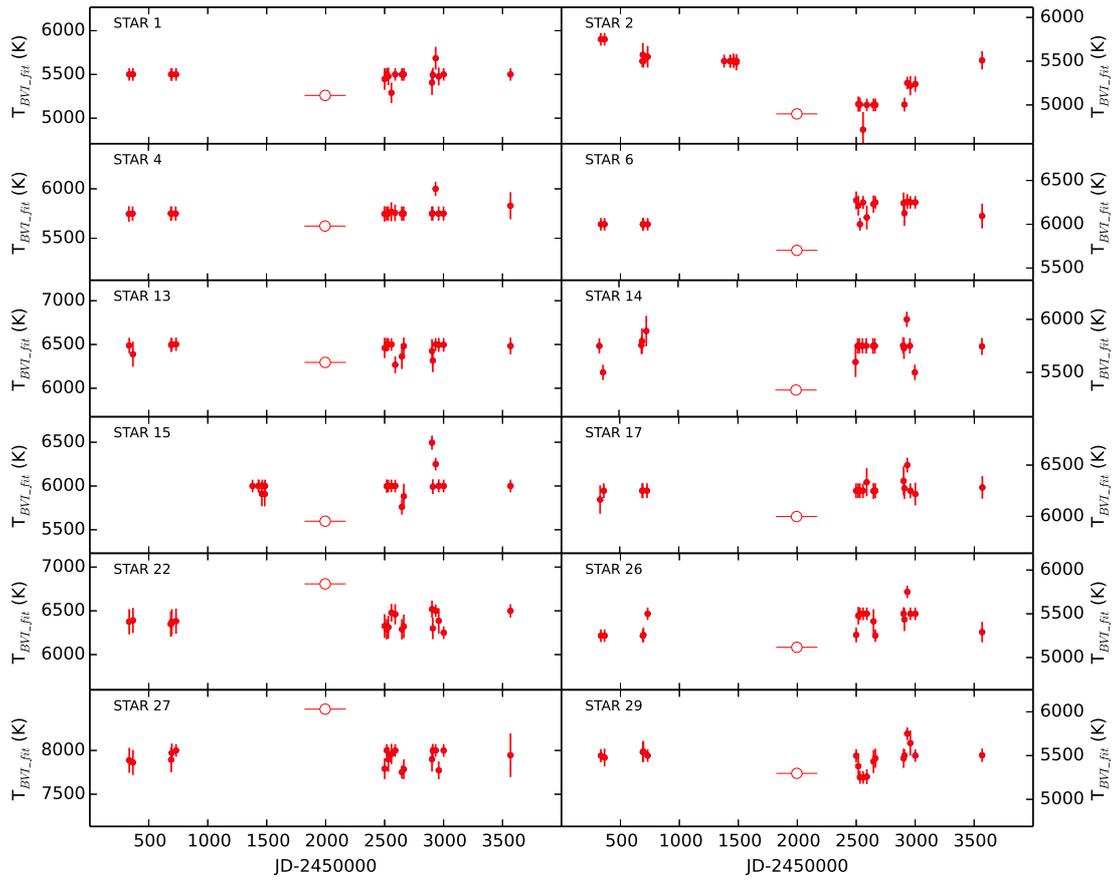}
\caption{Best-fit temperatures (filled circles) to the binned $BVI_{c}$ light curves, for the first 12 stars from Table \ref{tab05}. Open circles indicate photometric temperatures from \cite{Drout12}, derived using $B-V$ colors measured in northern fall 2000/2001 \citep{Massey06}. Plots of temperatures for the remaining nine stars are shown in the electronic edition of the current article.}
\label{fig07}
\end{figure*}

\begin{figure*} 
\centering
\includegraphics[width=7.5in]{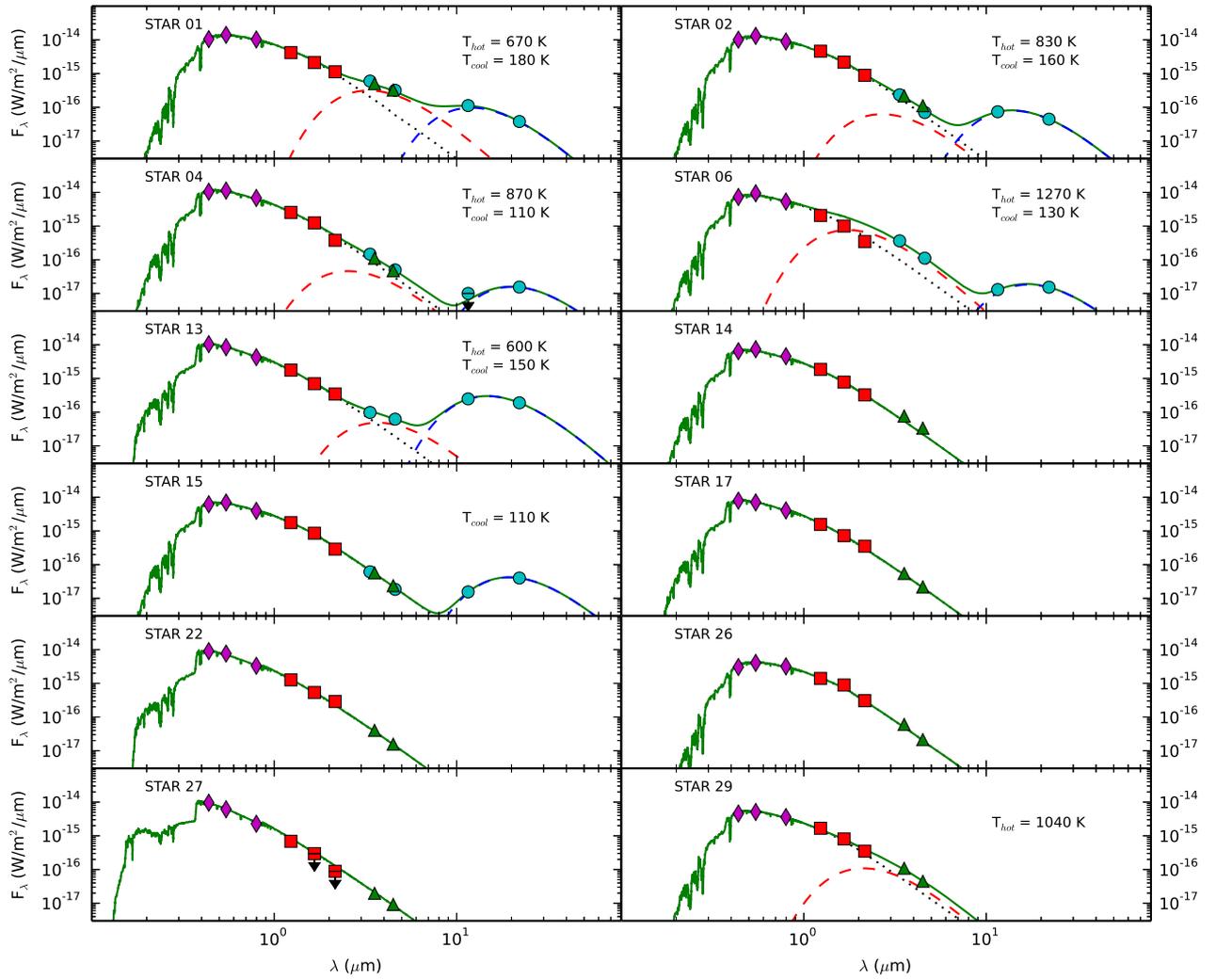}
\caption{Spectral energy distributions for the first 12 stars from Table \ref{tab05}. We show observations from DIRECT in $BVI_{c}$ (magenta diamonds), from 2MASS in $J$, $H$ , $K_{s}$ (red squares), from \textit{Spitzer} at [3.6] and [4.5] (green triangles) and from WISE at 3.4, 4.6, 12, and 22 $\mu$m (cyan circles). An arrow indicates an upper limit in the flux. The best-fitting model (solid green line) comprises the reddened, Kurucz photospheric component (dotted black line), a hot dust or free-free emission component (dashed red line) and a cool dust component (dashed blue line). Spectral energy distributions for the remaining nine stars are shown in the electronic edition of the current article.}
\label{fig08}
\end{figure*}

The SED-fit luminosities of the two most luminous YSGs, stars 1 and 2, appear to agree within errors to the luminosity estimates from the calibration of the \ion{O}{I} feature in the GTC spectra. This agreement, however, is marginal for star 1, given that the strength of \ion{O}{I} is additionally sensitive to the temperature for $T_{\text{eff}}<6\,000$ K. The rather strong \ion{O}{I} line of star 2 implies a higher temperature than that found by the SED fit (5\,500 K). The observed strength of the P14 line with respect to that of a 5\,500 K synthetic model also supports this conclusion (Fig. \ref{fig06}). A higher reddening value would obviously yield a higher SED temperature, however, later in this paper we infer a small value of the reddening from the optical photometry. Variability in the temperature between 1997 and 2014 would clearly justify this disagreement.

We find that eight out of 21 YSGs display evidence of hot dust with temperatures up to $\sim1\,200$ K, and eight stars show emission of cool dust with temperatures up to 180 K. Of the above, six YSGs display both hot and cool emission with the majority of them having luminosities log~$L/L_{\odot}\ge5.3$. Star 33 and possibly star 40 exhibit a NIR excess indicative of free-free emission from stellar winds. In addition, star 6 was matched with one 2MASS source within $6\arcsec$, but modeling with a blackbody does not allow to simultaneously fit both the near- and mid-IR fluxes. Because observations from the two surveys are separated in time, this could imply variability in the circumstellar environment of the star. Star 48 has a questionable dust excess and is designated with a `?'.

In addition to the 21 YSGs with time-series photometry, we similarly fit the SEDs for the three YSGs with spectra from GTC by employing $UBVRI$ counterparts from \cite{Massey06} and ATLAS9 models with log$g=0.5$. 

\begin{figure*} 
\centering
\includegraphics[width=5.5in]{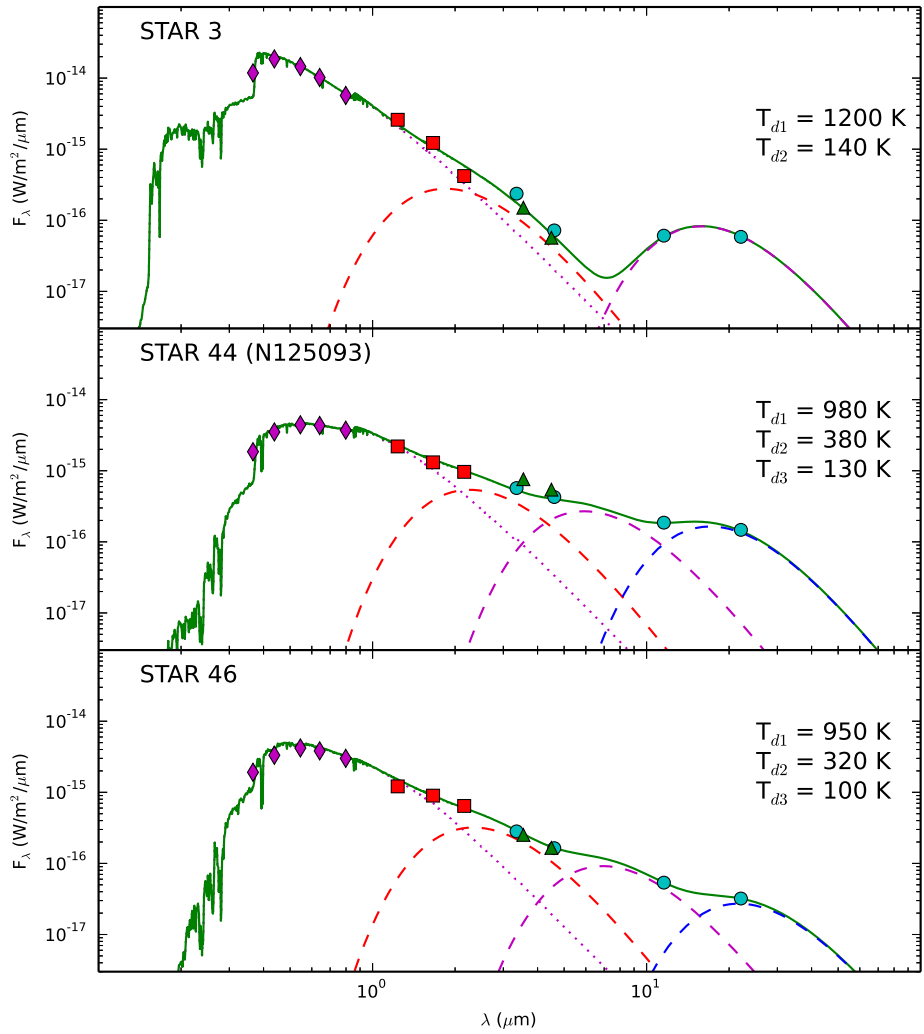}
\caption{Spectral energy distributions (similar to Fig. \ref{fig08}) for the three spectroscopic targets observed with GTC with no available time-series photometry. Optical $UBVRI$ counterparts are taken from \cite{Massey06}. An additional component (dashed magenta line) is used in two stars to fit thermal emission from warm dust.}
\label{fig09}
\end{figure*}

\begin{itemize}
\item{For the fit of N125093 (star 44) we adopted $A_{V}=1.6$ mag ($E(B-V)=0.5$ mag) measured by \cite{Humph13}. A spectrum with a temperature consistent to the SED-inferred one fits the observed P14 line well (Fig. \ref{fig06}), although, as previously stated, one should not rely on this agreement to infer spectroscopic $T_{\text{eff}}$ values.}

\item{The hotter star 3 shows a discrepancy between the MIR photometry and 2MASS, with the former indicating thermal emission from dust with $T=1\,200$ K. We found no second 2MASS or \textit{Spitzer} source within 6$\arcsec$ from the target and therefore, photometry likely originates from an infrared-variable source. The integrated luminosity of log~$L/L_{\odot}\sim5.6$ is consistent with the high luminosity indicated by the strength of \ion{O}{I}, and both deviate from the value derived from the $\ion{Ca}{II}$/Paschen region. As noted in Sect. \ref{spect}, filled-in emission of hydrogen could explain this discrepancy. At the same time, weakened hydrogen lines would result in an underestimation of the modeled temperature. Obviously, the SED-fit temperature of star 3 is hotter than the spectroscopic temperature of 7\,000 K, which could support this statement.}

\item{For star 46, a second source at $3.5\arcsec$ from the target is found in both 2MASS and \textit{Spitzer} databases. Since the IRAC [3.6] and [4.5] measurements for the object match these from WISE at the W1, W2 bands, uncertainty due to contamination would mainly concern the cool infrared excess.} 

\end{itemize}
The fitted SEDs of the three spectroscopic targets are shown in Fig. \ref{fig09}, with the corresponding parameters being listed in Table \ref{tab06}. Armed with spectroscopy, SED-fit parameters and, for stars 1 and 2, a long-term photometric monitoring, we discuss the evolutionary state of the most luminous targets in the following section.

\section{Discussion}
\label{disc}

\begin{figure*} 
\centering
\includegraphics[width=7.5in]{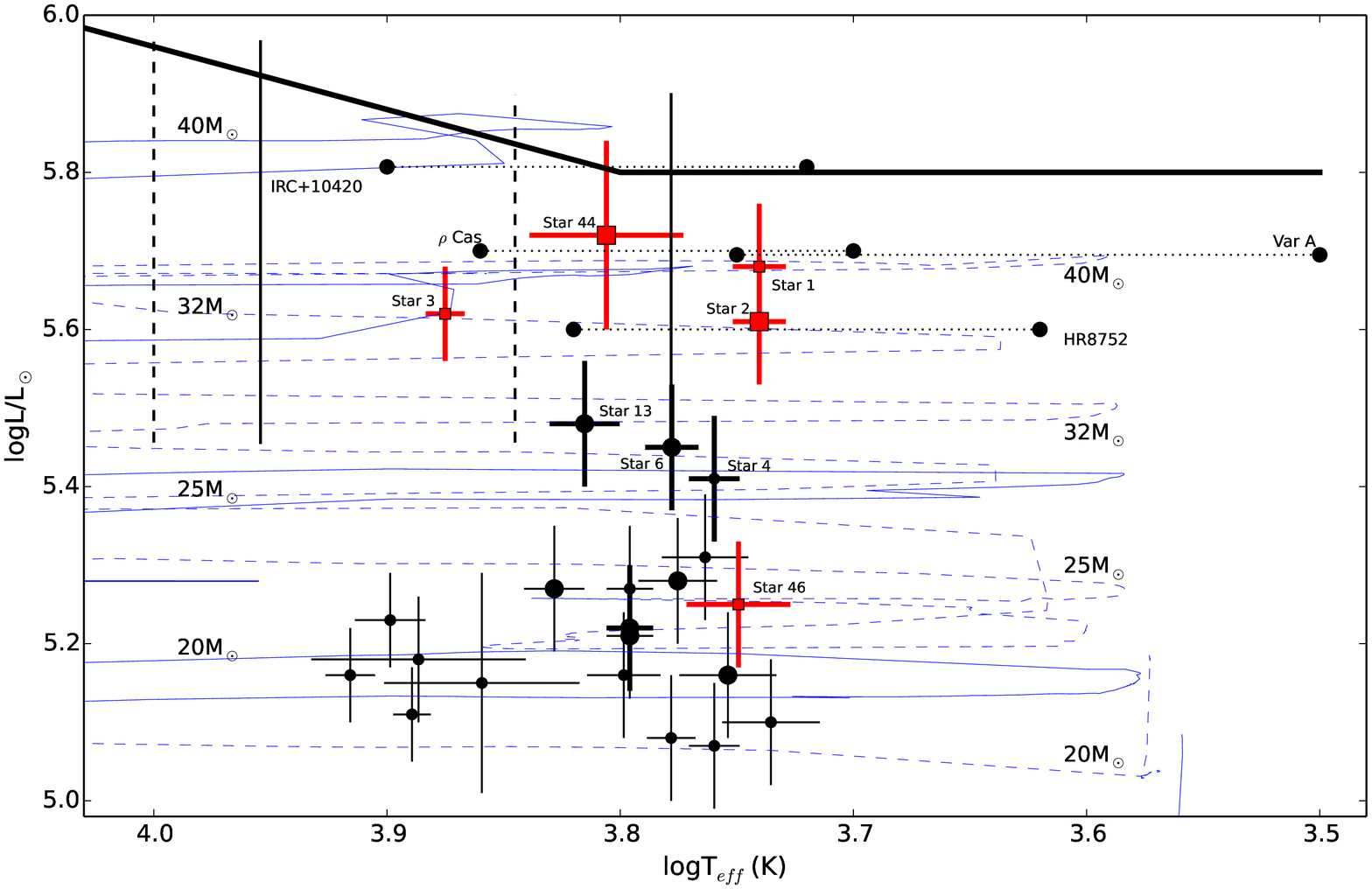}
\caption{Hertzsprung-Russell diagram for the 21 long-term monitored YSGs in M33 (black circles) and for the five spectroscopic targets (red squares). Error bars correspond to 2$\sigma$ uncertainties.  We emphasize with thick crosses those stars that show emission in the infrared from both hot and cool dust. The candidate YHGs/post-RSGs suggested in this study are marked with large symbols. Labels are shown for selected stars discussed in the text. The solid vertical lines at 6\,000 and 9\,000 K indicate the borders of the YV \citep{Humph02}, which extends above log~$L/L_{\odot}=5.4$ and below the Humphreys-Davidson limit (solid thick line). The suggested YV by \cite{deJ98}, lying within 7\,000$-$10\,000 K, is illustrated by the dashed vertical lines. Evolutionary tracks are taken from \cite{Ekstrom12} at $Z=Z_{\odot}$, non-rotating (dashed blue lines) and at the 40$\%$ of the critical velocity (solid blue lines) for masses indicated on the right for the former and on the left for the latter. The dotted lines connecting black solid circles indicate the variability in the temperature of four known YHGs: $\rho$ Cas \citep{Lob03}, IRC+10420 \citep{Oudm98}, HR8752 \citep{deJag97} in the Galaxy, and Var A \citep{Humph87} in M33.}
\label{fig10}
\end{figure*}

The SED-fit parameters of the 21 YSGs along with the three additional spectroscopic targets, stars 3, 44, and 46, are plotted on the HR diagram in Fig. \ref{fig10}. Error bars represent 2$\sigma$ uncertainties. We show the integrated luminosities at 22 $\mu$m because the additional energy output compared to the 8 $\mu$m-integrated energy is small with an exception for star 13. We show both borders of the YV proposed by \cite{Humph02} and \cite{deJ98} with luminosity log~$L/L_{\odot}>5.4$. Both sets of evolutionary tracks by \cite{Ekstrom12} at solar metallicity are shown, without rotation and assuming initial rotation at the $40\%$ of the critical velocity. Our studied sample displays initial masses of $20-40$ M$_{\odot}$ derived from the non-rotating models and $\sim18-35$ M$_{\odot}$ assuming rotation. 

Dust excess is not found for stars below log~$L/L_{\odot}\sim5.2$, which could imply a lower threshold for the post-RSG evolution in M33. Notably, this limit is also supported by the blueward turn of the 20 M$_{\odot}$ rotating model and is consistent with the results of \cite{Smart09}, which confine the RSG progenitors of Type II-P SNe below 17 M$_{\odot}$ and predict the blueward evolution of their more massive counterparts. A lack of YSGs with $T_{\text{eff}}>7\,000$ K above log~$L/L_{\odot}\sim5.35$ is evident, with the exception of star 3. It is interesting that, all seven stars of our sample, stars 1, 2, 3, 4, 6, 13 and 44, which occupy this upper region in the HR diagram, display both hot and cold dust emission. Being in agreement with the constraints set by the YV, these serve as potential YHGs. An eighth target, the less luminous spectroscopic star 46, displays a similar infrared profile. \cite{Gordon16} obtain moderate resolution spectroscopy for four of these targets, stars 1, 4, 44 (Sect. \ref{spect}), and 46. Although their observations do not cover the critical \ion{Ca}{II} near-infrared region, they provide sufficient resolution for the detection of forbidden lines such as [\ion{N}{II}] $\lambda6548,6583$ and [\ion{Ca}{II}] $\lambda7291,7323$ that emerge from gaseous extended envelopes and are frequently found in the optical spectrum of YHGs \citep[e.g.][]{Lob13,Ches14,Aret16}. By combining the available spectrophotometric data for each of the eight stars, we discuss their evolutionary state.

\begin{itemize}
\item{The DIRECT long-term photometry of star 1 indicates an invariant temperature in agreement with our spectroscopic one and therefore, we cannot claim variability typical of a YHG ``bounce'' in the last 20 years. We found emission of [\ion{N}{II}] and [\ion{S}{II}] along with narrow, single-component emission of H$\alpha$ in the spectrum of \cite{Gordon16}, which may be attributed to insufficient sky/nebular subtraction. No [\ion{Ca}{II}] is evident. Our absorption spectrum of star 1 implies that the star has not yet evolved as a YHG.}

\item{The location of star 3 on the HR diagram denotes that, if the star exhibited a post-RSG state, it would reside inside the YV. Our absorption spectrum, however, is not indicative of a star showing atmospheric instability and intense mass loss. We therefore find star 3 to be a luminous YSG.}

\item{Star 4 was not among our spectroscopic targets, however, the moderate resolution spectrum by \cite{Gordon16} shows weak [\ion{N}{II}] and [\ion{S}{II}] emission, lacks [\ion{Ca}{II}], and further displays hydrogen in absorption}, which is typically found in the spectrum of YSGs.

\item{Star 44 (N125093) was suggested by \cite{Valeev10} as a candidate LBV, based on their derived luminosity log~$L/L_{\odot}=6.3-6.6$. In the present study, we measure a luminosity log~$L/L_{\odot}=5.72\pm0.06$. The updated position of N125093 on the HR diagram favors the classification of \cite{Humph13} as a YHG candidate in accordance with both their and our spectroscopic diagnostics. The extended flat excess in the infrared regime of the star was fit with three dust components, which could be interpreted as dust layers formed during past mass loss events, which are expanding at different distances from the star. The spectrum by \cite{Gordon16} shows emission of [\ion{Ca}{II}] and [\ion{N}{II}], with the latter being blended with the wings of a strong and broad H$\alpha$ emission line.}

\item{We examined the possibility that star 46 exhibits an increased reddening caused by self-obscuration, which could result in an underestimation of the luminosity. The spectroscopic surveys of \cite{Massey16} and \cite{Gordon16} assign a spectral type to the star of G0Ia and F8, respectively, that in turn indicate an intrinsic $(B-V)_{0} = 0.69\pm0.14$ mag. Correction of the mean DIRECT $B-V$ value yields a low reddening of $E(B-V)=0.17\pm0.14$ mag. Applying this value to our analysis yields a luminosity of log~$L/L_{\odot}=5.28\pm0.07$ that is still below the limit to qualify as a YHG. In addition, no forbidden emission is shown in the spectrum of \cite{Gordon16}. We conclude that the star is likely affected by nebular emission from its surroundings, although the warm dust emission contributes a significant amount of energy above the photospheric component.}

\end{itemize}
We searched the literature for evidence that would aid us in our classification of the unexplored stars 2, 6, and 13 as YHGs. We found that all three stars were characterized by \cite{Shpo06} as variable sources from their $V-$band monitoring of M33 during the period $2000-2003$. We added their published light curves to our DIRECT ones, having first corrected for an offset of $0.25$ mag as indicated by \cite{Shpo06} for stars of $V=16-16.5$ mag. For star 6, we report a periodic variability of approximately 400 days with an amplitude measured to be roughly 0.4 mag. A lower-amplitude variability of $\sim0.1$ mag is also shown for star 13, although periodicity is not as clear. Both stars show fluctuations in the photometric temperature that do not exceed 250 K (Fig. \ref{fig07}), which may indicate pulsational activity. Whether the observed modulation is associated with the physics of a post-RSG state requires future spectroscopic investigation and hence, we classify these stars as YHG candidates. On the other hand, star 2 exhibits significant variability in both the $V-$band light curve and the photometric temperature, which is further supported by the optical spectroscopy presented in the current study.

The less luminous stars 15, 29, 31, 32, and 33, show infrared excess emerging from one or two dust components, with star 33 displaying free-free emission in the NIR due to winds. A moderate resolution study is only available for star 15, which interestingly shows blueshifted emission of [\ion{O}{I}]. We classify these stars as post-RSGs candidates to be further explored by follow-up spectroscopy.

\subsection{Star 2}
\label{star2}

\begin{figure*} 
\centering
\includegraphics[width=5.3in]{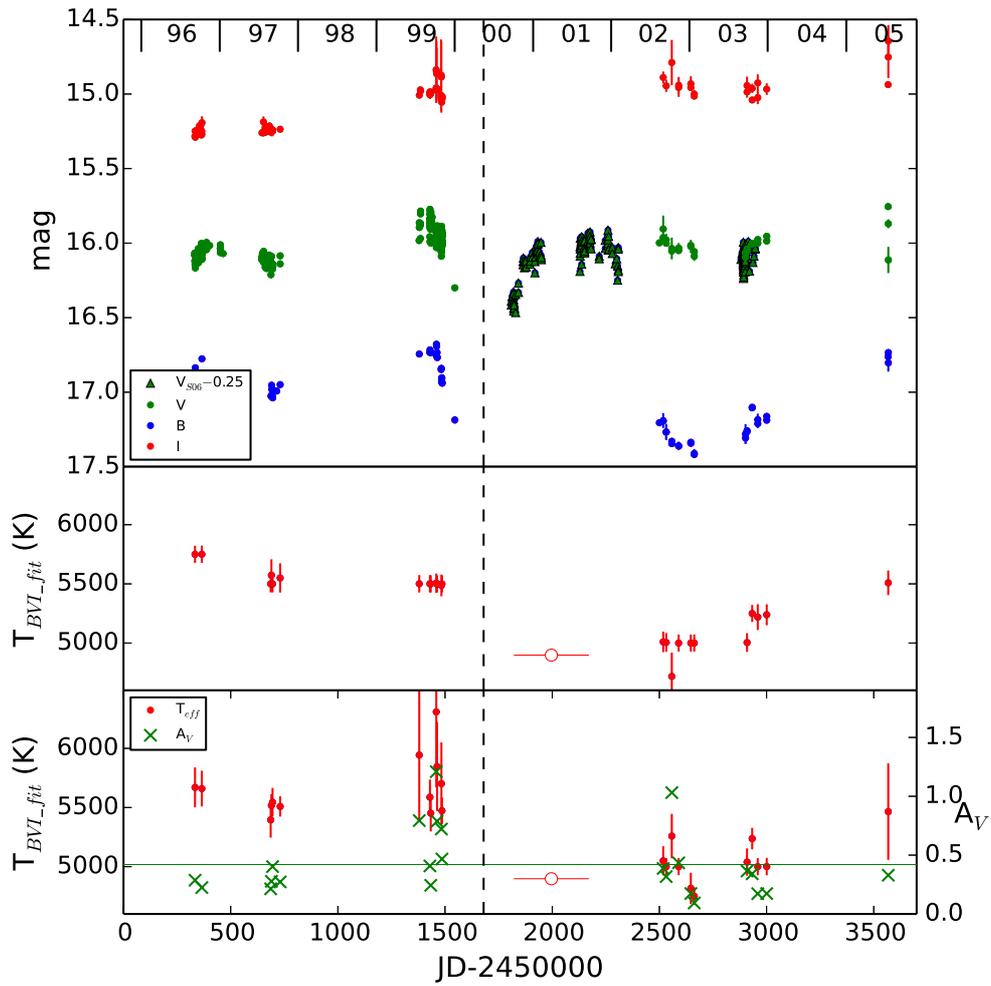}
\caption{\textit{Upper panel.} The $BVI_{c}$ light curves for star 2 \citep{Macri01,Pelle11}. $V-$band photometry from \cite{Shpo06} is shown as green triangles, corrected by 0.25 mag. Uncertainties are indicated with error bars. The dashed line at JD 2\,451\,680 (mid-May 2000), serves as an approximation for the date when the fading event reaches a minimum. \textit{Middle panel.} A zoomed view of the best-fit temperatures for star 2 from Fig. \ref{fig07}. \textit{Lower panel.} Same as in the middle panel, but having visual extinction (green crosses) as a free parameter. A green line indicates the average extinction of M33, at $A_{V}=0.4$ mag. }
\label{fig11}
\end{figure*}

\begin{figure*} 
\centering
\includegraphics[width=5in]{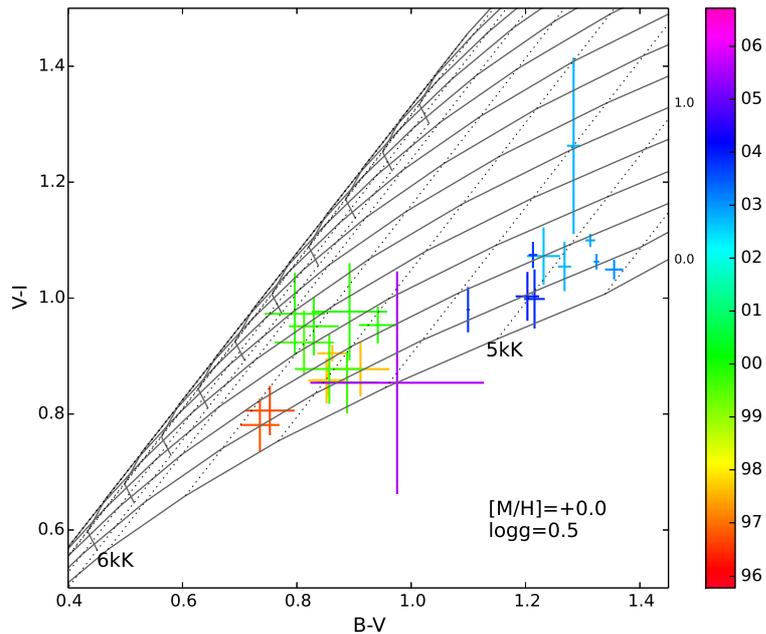}
\caption{Time evolution of star 2 on the $V-I$ vs. $B-V$ CCD, color coded by year, superposed on a grid of $T_{\text{eff}}$ and $A_{V}$. From right to left, temperature increases with a step of 250 K and is shown as a dotted line. Temperatures of 5kK and 6kK are noted. From the bottom up, the solid lines indicate models with increasing extinction, with a step of 0.2 mag. $A_{V}$ values of 0.0 and 1.0 mag are noted in the right side of the panel. We note the degeneracy of the parameters for $T_{\text{eff}}>6\,000$ K.}
\label{fig12}
\end{figure*}

The light curves of star 2 in $B$, $V$ and $I_{c}$ from DIRECT along with the $V$-band light curve from \cite{Shpo06} are shown in the upper panel of Fig. \ref{fig11}. Uncertainties are on the order of 0.01 mag. In the middle panel, the fit to the DIRECT photometry with synthetic models indicates a decrease in temperature after JD 2\,451\,500, which at first glance, has an amplitude of 750 K compared to the temperatures derived by measurements at the beginning of the monitoring. The binned $BVI_{c}$ photometry and the derived $T_{\text{eff}}$ values are listed in Table \ref{tab07}. Our analysis assumed a reddening fixed to the average value of M33, given that modeling above 6\,000 K is constrained by the degeneracy between $T_{\text{eff}}$ and $A_{V}$ when proceeding with the $B-V$ and $V-I_{c}$ colors. In the temperature range $4\,750-6\,000$ K of star 2 however, both $T_{\text{eff}}$ and $A_{V}$ are well determined by the observed colors. To illustrate this, we built a grid of the effective temperature of the Kurucz models and the visual extinction with a step of $\Delta A_{V}=0.2$ mag as a function of the two colors, as shown on the CCD in Fig. \ref{fig12}. The synthetic $B-V$, $V-I_{c}$ colors were defined for each reddened Kurucz model using the zeropoints for the respective color indices provided by \cite{Bess98}. The binned color data of star 2 are superimposed on the grid and are color coded as a function of time. As noted, the degeneracy between $T_{\text{eff}}$ and $A_{V}$ for temperatures above 6\,000 K is strongly evident, and the DIRECT observations are well confined below 6\,000 K.

We therefore repeated the fitting process described in Sect. \ref{temp}, with $A_{V}$ as a free parameter. The results are shown in the lower panel of Fig. \ref{fig11} and are listed in Table \ref{tab07}. Compared to the temperatures derived with a fixed $A_{V}$, an increase in the $T_{\text{eff}}$ is now shown around JD 2\,451\,400 followed by an increase in the $A_{V}$. Uncertainties in the estimated values in this observing window are clearly larger, which is partly attributed to the uncertainty in the averaged photometric data, but could also imply an increase of temperature to around and above 6\,000 K where parameters are not well-constrained due to the aforementioned degeneracy. On average, the new extinction values for star 2 are slightly lower than the average extinction of M33 but similar to the values found for other supergiants in M33 (Table \ref{tab05}). We find a good agreement between the parameters derived from the $\chi^{2}$-minimization and those inferred from interpolating the binned color data on the grid of parameters in Fig. \ref{fig12}. 

The decrease in the temperature of star 2 appears to follow a notable event shown in the middle of the observing coverage. The suppression of the visual light curve from \cite{Shpo06} around JD 2\,451\,800 indicates a large drop that exceeds 0.5 mag in depth. The faint point obtained on JD 2\,451\,545 in both $B$ and $V$ for DIRECT helps to roughly constrain the minimum of the event to around JD 2\,451\,675, mid-May 2000. We outline some remarkable similarities to the light curve of the well-studied YHG $\rho$ Cas, which was thoroughly investigated by \cite{Lob03}.

In late 2000, the YHG $\rho$ Cas exhibited a dimming of $1.2-1.4$ mag following the so-called millenium outburst that caused an apparent drop in the temperature of $\sim3\,000$ K. During the drop, high resolution spectroscopy revealed the formation of a cool TiO envelope, which expanded faster than the photosphere and lasted for almost half a year. The increase in the $V-$band brightness in $\rho$ Cas prior to the dimming indicates the heating of the atmosphere at $\sim8\,000$ K, which is the temperature for the partial ionization of hydrogen. Driven by the energy of recombination during the expansion, the outburst resulted in a high mass-loss rate of $\sim5\times10^{-2}$ M$_\odot$ yr$^{-1}$. In addition to the dimming, the light curve of $\rho$ Cas displays fluctuations that are attributed to photospheric pulsations and changes in the temperature of $500-800$~K. Star 2 similarly exhibited a large drop in the $V-$band almost six months before the dimming of $\rho$ Cas that coincides with the evident decrease in temperature. The composite visual light curve of star 2 indicates additional dips around JD 2\,452\,300 and 2\,452\,890 with a duration of less than 200 days that may correspond to pulsational activity similar to that of $\rho$ Cas. Prior to the large decline, star 2 reached a visual maximum of 15.9 mag, which is associated with an increase in the value of $A_{V}$, implying a high mass-loss episode similar to the outburst of $\rho$ Cas. We caution that our analysis could have underestimated the parameters during the outburst, owing to the inability of the Kurucz LTE models to account for the atmospheric instability and the possible non-spherical mass loss of a YHG when it approaches the YV.

Unfortunately, we lack photometric observations during the minimum of star 2 that would specify the coolest epoch of the star. Almost three years after the DIRECT measurements of 1999, star 2 was cooler by at least 500 K. As previously noted, compared to a 5\,500 K model for the pre-outburst state of 1997, the GTC spectrum from 2014 shows a stronger hydrogen feature at $\sim8598$ $\AA$ (Fig. \ref{fig06}) that is in line with the increase in the temperature shown at the end of the photometric monitoring. Had the star exhibited a so-called ``bounce'', this gradual heating would be justified by the recovery of the star back to its warm evolutionary state, having exhibited a large change in its atmospheric thermal structure or in the opacity of the wind. In the former case and assuming a constant luminosity, the observed change in temperature implies an expansion of the stellar radius by approximately $48\%$. For comparison, $\rho$ Cas showed a displacement in the radius by approximately $25\%$ over a single pulsational period in 1970 \citep{Lob94} and by a factor of two during the outburst of 2000 \citep{Lob03}.

We classify star 2 as a strong YHG candidate based on the spectroscopic signs of an extented atmospheric structure, a dust excess emerging in the infrared regime and an enigmatic color variability that suggests instability of the outer layers of the star. The emission of \ion{Ca}{II} is also observed in the spectrum of Galactic ``hot'' YHGs that approach the instability border of the YV, and has been attributed to the presence of Keplerian rotating disks that were formed by material expelled in the equatorial plane \citep{Aret16}. This scenario would support our conclusion for star 2 being a YHG located near the YV and would further suggest the pre-RSG state of the hot star 3 showing \ion{Ca}{II} in absorption. We emphasize, however, the need of high-resolution spectroscopy for an accurate YHG classification, which relies on measuring the intrinsic broadening of the absorption photospheric lines according to the Smolinski criterion.

\subsection{B324}
\label{b324}

In addition to the study of YSGs, we undertook $K-$band spectroscopy of the extreme warm super/hypergiant B324 (\object{LGGS J013355.96+304530.6}) in M33 with IRCS/Subaru. Interestingly, B324 is located close to star 2 at a projected distance of 30$\arcsec$ (Fig. \ref{fig02}), corresponding to 140 pc. The true evolutionary state of B324 is controversial with arguments for both a YHG and an high-luminous LBV in eruption given in the literature because of uncertainty in the distance \citep{Clark12,Humph13}. Its $K-$band spectrum is shown in Fig. \ref{fig04}. Typically, the CO band head profile at $2.3-2.4$ $\mu$m is shown in absorption for cool supergiants with $T_{\text{eff}}<5\,000$ K. Higher temperatures cause the dissipation of the molecule, except for the cases when it is concentrated in disk-like structures as observed in supergiant B[e] stars and several YHGs that show emission of the feature \citep{Oksa13}. Additionally, a small ratio ($\le15$) of $^{12}$CO/$^{13}$CO is anticipated for evolved stars resulting from an $^{13}$CO enrichment on the surface \citep{Kraus09}. The variable CO profile of $\rho$ Cas, however, is found to originate close to the stellar photosphere and displays a strong dependence on the pulsational cycles, being in emission only at phases of maximum brightness \citep{Gorlo06}. Consequently, an absence of the molecule in emission could be attributed to the geometrical distribution of the expelled material and thus would not prevent a YHG classification. For B324, the absence of CO in absorption is consistent with the warm temperature of the star at 8\,000~K \citep{Humph13} and blueshifted hydrogen is present in weak emission indicating mass loss. The most prominent feature is the strong emission of \ion{Na}{I} at $\sim2.2$ $\mu$m that is also shown in the near-infrared spectra of the YHGs $\rho$ Cas, HR8752 and most prominently of IRC+10420, and is suggested as an indicator for the overabundance of Na \citep{Yama07}. Similarities between the optical spectra of B324 and IRC+10420 are also pointed out by \cite{Humph13}.

\section{Conclusions}
\label{concl}

We have used the catalog of YSGs in M33 from \cite{Drout12} to investigate the poorly-studied class of YHGs beyond the Galaxy. We conducted optical spectroscopy of five luminous YSGs showing evidence of surrounding dust and additionally focused on the time-variant properties of the 21 most luminous YSGs. We expect variations in their spectral type on timescales of years or decades because YHGs undergo dramatic atmospheric changes as they ``bounce against the YV''. We made use of atmospheric models to fit the $BVI_{c}$ long-term photometry available for the 21 YSGs in an attempt to trace high-amplitude fluctuations in the apparent effective temperature. 

We find that the upper part of the HR diagram (log~$L/L_{\odot}\gtrsim5.35$) is occupied by seven YSGs, stars 1, 2, 3, 4, 6, 13, and 44, which show infrared excess that arises from cold and hot dust emission. Of these stars, we classify the photometric variables stars 2, 6, and 13 as YHG candidates, to be confirmed by follow-up high-resolution spectroscopy. We find the luminous stars 1, 3 and 4 to lack evidence typically shown in the spectra of YHGs and therefore, they are likely to be pre-RSG stars. The less-luminous, dusty YSGs stars 15, 29, 31, 32, and 33 are proposed as candidate post-RSG stars.

Star 2 constitutes a rather promising YHG candidate supported by long-term variability that partly resembles the ``bounce'' of a YHG against the cool border of the YV. We report a decrease in temperature of more than 500 K that followed a dimming of amplitude larger than 0.5 mag in the $V-$band light curve of the star. We attribute the observed variability to a modification in the atmospheric structure and/or in the wind optical depth that followed a possible outburst that occurred in mid 1999. In addition, our optical low-resolution spectroscopy of the object displays more than one H$\alpha$ emission components and emission of [\ion{N}{II}] and \ion{Ca}{II} as a result of an expanding gaseous shell. Emission from dust in the infrared implies that the star may have undergone several eruptive mass-loss events in the past. For N125093/star 44, we provide a temperature of $6\,400\pm240$ K and luminosity log~$L/L_{\odot}\sim5.72\pm0.06$ that renders the star a candidate YHG because it is rather too cool to be a LBV in eruption as previously suggested. Additionally, N125093 displays significant infrared excess fit with three dust components in the range $\sim100-900$ K that is indicative of past mass-loss events. Near-infrared spectroscopy of B324 revealed emission of Na also shown in the $K-$band spectra of other YHGs, suggesting the post-RSG nature of the star instead of a highly-luminous LBV.

Our method to combine time-series photometry with optical spectroscopy yielded a strong YHG candidate, star 2, which if confirmed, will be the second extragalactic YHG to exhibit a ``bounce'' after Var A. We suggest measurements of the intrinsic broadening of the photosperic lines and the identification of weak forbidden emission via high-resolution spectroscopy to confirm the YHG nature of star 2 and the other YHG candidates. Future work will include the study of the remaining luminous YSGs in M33 from \cite{Drout12} that were confirmed spectroscopically by \cite{Massey16}. 

Long-term multi-band photometric monitoring of YSGs in other galaxies of the Local Group will contribute to populating the warm hypergiant region of the HR diagram and therefore provide significant constraints to the theoretical framework of YHGs as a function of metallicity. The deep multiband time-series photometry from the Panoramic Survey Telescope And Rapid Response System \citep[Pan-STARRS;][]{Chamb16} will allow the detection of variability in the spectral type of distant and/or dust-enshrouded yellow stars and provide a firm database of critical targets for future spectroscopic observations with the next generation of telescopes such as the James Webb Space Telescope (JWST) and the European-Extremely Large Telescope (E-ELT).

\begin{acknowledgements}
We thank Drs. R. Humphreys and M. Gordon for their helpful comments on an earlier draft of the manuscript, and the anonymous referee for comments that improved this work. M. Kourniotis and A.Z. Bonanos acknowledge  funding  by  the  European  Union  (European  Social Fund) and National Resources under the ``ARISTEIA'' action of the Operational Programme ``Education and Lifelong Learning'' in Greece. This research has made use of NASA’s Astrophysics Data System Bibliographic Services and the VizieR catalogue access tool, CDS, Strasbourg, France. 
\end{acknowledgements}

\clearpage

\clearpage

\begin{sidewaystable*}
\caption[]{Infrared photometry of 53 luminous YSGs in M33.}
\label{tab01}  

{\tiny
{\centering
\begin{tabular}{ccccccccccccccc}

\hline \hline
ID & LGGS  &  $J$  &  $H$  &$K_{s}$&  $[3.6]$  &  $[4.5]$ &   W1  &    W2  &    W3   &   W4  &  Code\footnote{Within 6$\arcsec$ from the target, \textbf{(0)} one identified source by 2MASS and/or \textit{Spitzer} \textbf{(1)} at least two identified sources by 2MASS or \textit{Spitzer} \textbf{(2)} at least two identified sources by both 2MASS\\ and \textit{Spitzer}.} & \multicolumn{2}{c}{Sp. Type\footnote{(M16) \cite{Massey16}, (G16) \cite{Gordon16}}} & Note\footnote{(M01) Photometry by \cite{Macri01}, (GTC) Optical spectroscopy with OSIRIS/GTC, (SUB) $K-$band spectroscopy with IRCS/Subaru, (S06) Variable in \cite{Shpo06},\\ (IR-VAR) Variable in at least one IRAC band in \cite{Quin07}.} \\ 
   &       & (mag) & (mag) & (mag) &   (mag)   &   (mag)  & (mag) &  (mag) &  (mag)  & (mag) &        &   M16       & G16   &   \\
\hline
1 & J013349.86+303246.1 & 14.68 (0.04) &14.30 (0.05) & 13.95 (0.04) &  12.81 (0.05) & 12.29 (0.06) &12.85 (0.04) & 12.21 (0.03) &  9.50 (0.04) & 7.83 (0.16)&  2  & G0Ia &  F8  & M01, GTC, S06 \\         
2 & J013358.05+304539.9 & 14.58 (0.03) &14.27 (0.05) & 14.21 (0.05) &  13.72 (0.05) & 13.45 (0.05) &13.86 (0.04) & 13.85 (0.05) &  9.96 (0.08) & 7.66 (0.11)&  2  &      &      & M01, GTC, SUB, S06 \\    
3 & J013411.32+304631.4 & 15.21 (0.04) &14.90 (0.07) & 15.03 (0.12) &  14.12 (0.09) & 14.18 (0.10) &13.86 (0.04) & 13.82 (0.05) & 10.17 (0.07) & 7.36 (0.08)&  0  &      &      & GTC, S06 \\             
4 & J013337.31+303111.1 & 15.23 (0.03) &14.88 (0.06) & 15.14 (0.11) &  14.45 (0.06) & 14.38 (0.06) &14.38 (0.04) & 14.22 (0.07) & 12.13 (U)  & 8.81 (0.31)&  1  & G0Ia &  F8  & M01 \\                   
5 & J013229.20+303430.2 & 15.12 (0.04) &14.94 (0.07) & 14.78 (0.08) &        $-$    &     $-$      &      $-$    &      $-$     &      $-$     &     $-$    &     & A8Ip &   A8  & \\                       
6 & J013351.84+303827.4 & 15.46 (0.05) &15.12 (0.09) & 15.23 (0.14) &        $-$    &     $-$      &13.39 (0.04) & 13.35 (0.05) & 11.84 (U)  & 8.82 (U) &  0  &      &      & M01, S06 \\              
7 & J013337.04+303637.6 & 15.27 (0.04) &15.14 (0.08) & 14.96 (0.12) &        $-$    &     $-$      &      $-$    &      $-$     &      $-$     &     $-$    &     &      & F5-8 & S06 \\              
8 & J013233.85+302728.9 & 15.21 (0.06) &14.68 (0.06) & 14.39 (0.07) &        $-$    &     $-$      &13.81 (0.03) & 13.67 (0.03) & 10.44 (0.06) & 7.54 (0.12)&  0  & F8I  &  F2  & \\                       
9 & J013327.65+303751.2 & 16.19 (0.09) &15.90 (0.16) & 15.80 (U)  &  15.60 (0.07) & 15.61 (0.06) &      $-$    &      $-$     &      $-$     &     $-$    &     &      & A5-8 & \\                   
10 & J013420.95+303039.9 & 15.99 (0.07) &15.89 (0.15) & 15.85 (0.21) &  15.46 (0.07) & 15.35 (0.08) &15.41 (0.05) & 14.95 (0.07) & 11.29 (0.10) & 8.43 (0.19)&  0  & A8Ia & A5-8 & \\                       
11 & J013311.16+303421.8 & 15.91 (0.06) &15.69 (0.12) & 15.63 (0.19) &  15.37 (0.05) & 15.30 (0.06) &15.46 (0.08) & 15.11 (0.09) & 10.96 (0.08) & 9.45 (0.50)&  1  & F0Ia &      & S06\\                   
12 & J013300.77+303416.9 & 16.37 (0.10) &14.76 ( $-$ )& 14.71 (U)  &  14.47 (0.06) & 14.35 (0.07) &14.11 (0.03) & 13.80 (0.04) &  9.86 (0.05) & 7.01 (0.08)&  2  & G0Ia &  F8  & \\                       
13 & J013345.15+303620.1 & 15.62 (0.07) &15.50 (0.12) & 15.24 (0.14) &        $-$    &     $-$      &14.81 (0.33) & 13.98 (0.14) &  8.64 (0.08) & 6.08 (0.06)&  0  & G0Ia &      & M01, S06\\                   
14 & J013352.56+303815.9 & 15.28 (0.07) &14.46 (0.07) & 14.00 (0.05) &  14.85 (0.11) & 14.75 (0.09) &12.23 (0.03) & 12.34 (0.03) & 11.84 (U)  & 9.02 (U) &  2  &      &      & M01, S06 \\              
15 & J013357.70+304824.1 & 15.62 (0.06) &15.28 (0.09) & 15.43 (0.16) &  15.17 (0.06) & 15.13 (0.06) &15.32 (0.07) & 15.30 (0.11) & 11.65 (0.38) & 7.77 (0.15)&  1  &      &  F5  & M01, S06 \\              
16 & J013439.98+303839.4 & 15.67 (0.06) &15.36 (0.10) & 15.20 (0.13) &  15.31 (0.05) & 15.22 (0.08) &15.01 (0.04) & 14.71 (0.06) & 11.57 (0.13) & 9.16 (U) &  1  & G0Ia &  F5  & \\                       
17 & J013351.65+303225.9 & 15.76 (0.06) &15.47 (0.10) & 15.21 (0.13) &  15.22 (0.06) & 15.22 (0.06) &      $-$    &      $-$     &      $-$     &     $-$    &     &      &      & M01, S06 \\              
18 & J013351.48+305252.9 & 15.88 (0.07) &15.64 (0.14) & 15.59 (0.18) &  15.39 (0.06) & 15.43 (0.07) &      $-$    &      $-$     &      $-$     &     $-$    &     &      &  F0  & \\                       
19 & J013431.87+304118.0 & 16.20 (0.09) &16.23 (0.20) & 16.14 (0.30) &  15.71 (0.05) & 15.70 (0.06) &15.86 (0.10) & 15.24 (0.10) & 10.99 (0.09) & 8.90 (0.31)&  1  & A8Ia &      & \\                       
20 & J013420.34+304545.6 & 15.54 (0.06) &15.15 (0.08) & 15.21 (0.13) &  14.59 (0.07) & 14.53 (0.06) &14.45 (0.25) & 14.05 (0.22) & 10.17 (0.18) & 9.13 (0.37)&  1  &      & F8-G0 & S06 \\                  
21 & J013349.56+303941.6 & 15.84 (0.10) &15.76 (0.16) & 15.21 (0.14) &      $-$      &     $-$      &      $-$    &      $-$     &      $-$     &     $-$    &     &      &      & \\                       
22 & J013348.36+304242.1 & 15.98 (0.07) &15.80 (0.16) & 15.42 (0.16) &  15.55 (0.10) & 15.58 (0.09) &      $-$    &      $-$     &      $-$     &     $-$    &     &      &      & M01, S06 \\              
23 & J013336.61+302208.1 & 16.12 (0.08) &15.98 (0.16) & 15.79 (0.20) &  15.59 (0.05) & 15.56 (0.06) &15.63 (0.05) & 15.22 (0.09) & 11.40 (0.12) & 8.74 (0.26)&  1  &      &  F0  & \\                       
24 & J013355.47+310009.0 & 15.98 (0.07) &15.52 (0.12) & 15.20 (0.12) &      $-$      &     $-$      &15.40 (0.05) & 15.11 (0.09) & 12.94 (U)  & 9.15 (U) &  0  &      &  F2  & \\                       
25 & J013311.09+304851.8 & 15.52 (0.05) &15.21 (0.09) & 15.10 (0.12) &  14.79 (0.10) & 14.84 (0.09) &14.35 (0.03) & 14.29 (0.05) & 11.51 (0.16) & 8.43 (0.22)&  1  &      &  F8  & \\                       
26 & J013357.48+303821.4 & 15.88 (0.07) &15.24 (0.10) & 15.37 (0.15) &  15.11 (0.05) & 15.24 (0.08) &13.91 (0.07) & 13.84 (0.10) & 11.23 (U)  & 9.05 (U) &  2  &      &      & M01, S06 \\              
27 & J013356.58+303826.6 & 16.65 (0.13) &16.44 ( $-$ )& 16.71 (U)  &  16.33 (0.12) & 16.16 (0.12) &      $-$    &      $-$     &      $-$     &     $-$    &     &      &      & M01, S06 \\                  
28 & J013419.52+304633.8 & 16.31 (0.11) &16.05 (0.19) & 15.65 (U)  &  15.83 (0.05) & 15.82 (0.06) &      $-$    &      $-$     &      $-$     &     $-$    &     &      &      & \\                       
29 & J013344.15+303205.7 & 15.68 (0.06) &15.35 (0.10) & 15.22 (0.14) &  14.47 (0.11) & 14.44 (0.11) &13.79 (0.04) & 13.49 (0.05) & 8.66	(0.03)  & 6.80 (0.07)&  1  &      &      & M01, S06 \\              
30 & J013419.51+305532.4 & 16.18 (0.09) &15.79 (0.14) & 15.67 (U)  &  15.70 (0.06) & 15.87 (0.06) &      $-$    &      $-$     &      $-$     &     $-$    &     &      & A8-F0 & \\                       
31 & J013355.62+303500.8 & 15.80 (0.06) &15.58 (0.12) & 15.59 (0.18) &  14.80 (0.11) & 14.92 (0.09) &      $-$    &      $-$     &      $-$     &     $-$    &     &      &      & M01, S06 \\              
32 & J013410.90+303840.8 & 15.92 (0.07) &15.42 (0.10) & 15.59 (0.18) &  14.83 (0.10) & 14.92 (0.08) &14.71 (0.07) & 15.47 (0.15) & 11.33 (0.24) & 9.07 (U) &  0  & G0Ia &      & M01, S06 \\              
33 & J013348.48+304510.6 & 15.64 (0.06) &15.13 (0.08) & 14.86 (0.10) &      $-$      &     $-$      &14.89 (0.30) & 14.87 (0.22) & 10.94 (0.14) & 8.91 (0.36)&  0  & F5Ia &      & M01 \\                   
34 & J013322.62+302920.4 & 15.85 (0.06) &15.32 (0.09) & 14.98 (0.11) &  15.05 (0.09) & 14.89 (0.06) &14.77 (0.04) & 14.56 (0.06) & 11.44 (0.12) & 9.20 (U) &  1  &      & F8-G0 & \\                       
35 & J013303.60+302903.4 & 15.69 (0.05) &15.39 (0.09) & 15.22 (0.14) &  15.02 (0.05) & 14.81 (0.06) &14.85 (0.05) & 14.39 (0.05) & 10.67 (0.06) & 8.03 (0.15)&  0  &      & F8-G0 & IR-VAR \\                    
36 & J013358.51+303412.6 & 15.84 (0.08) &14.58 ( $-$ )& 14.34 (U)  &  15.27 (0.09) & 15.37 (0.09) &      $-$    &      $-$     &      $-$     &     $-$    &     &      &      & M01\\                       
37 & J013418.54+303647.5 & 15.84 (0.07) &15.51 (0.12) & 15.74 (0.21) &  15.33 (0.05) & 15.35 (0.06) &15.35 (0.11) & 16.53 (0.38) & 12.27 (U)  & 8.98 (U) &  0  & G0Ia &  F8  & \\                       
38 & J013342.84+303835.6 & 16.12 (0.10) &16.02 (0.19) & 16.04 (0.32) &  15.29 (0.12) & 15.46 (0.08) &      $-$    &      $-$     &      $-$     &     $-$    &     & G0Ia &      & M01 \\                   
39 & J013350.71+304254.3 & 16.69 (0.14) &16.12 (0.21) & 16.49 (U)  &      $-$      &     $-$      &      $-$    &      $-$     &      $-$     &     $-$    &     & A6Ia &      & M01, S06\\                   
40 & J013346.16+303448.5 & 16.31 (0.10) &15.89 (0.16) & 15.81 (0.23) &  15.58 (0.06) & 15.67 (0.06) &14.56 (0.08) & 14.99 (0.13) & 11.89 (U)  & 9.39 (U) &  2  &      &      & M01 \\                    
41 & J013320.08+302544.4 & 16.06 (0.08) &15.75 (0.16) & 15.99 (0.27) &  15.38 (0.05) & 15.38 (0.06) &14.91 (0.05) & 14.70 (0.07) & 10.68 (0.06) & 8.79 (0.27)&  1  &      &  F0  & \\                       
42 & J013252.56+303419.6 &   $-$          &  $-$      &   $-$        &  16.46 (0.06) & 16.46 (0.07) &      $-$    &      $-$     &      $-$     &     $-$    &     &      &  A8  & G16\\                   
43 & J013440.51+303852.3 & 16.39 (0.11) &16.64 (0.30) & 17.04 ( $-$) &  16.01 (0.05) & 16.02 (0.06) &      $-$    &      $-$     &      $-$     &     $-$    &     &      &      &\\                       
44 & J013415.42+302816.4 & 15.39 (0.04) &14.82 (0.06) & 14.12 (0.05) &  12.36 (0.05) & 11.73 (0.06) &12.90 (0.03) & 11.89 (0.02) & 8.95 (0.02)  & 6.36 (0.05)&  0  & F0Ie & F0-2 & GTC\\                   
45 & J013304.66+303240.4 & 16.59 (0.12) &16.44 (0.27) & 15.09 (U)  &  16.03 (0.05) & 15.96 (0.06) &14.21 (0.04) & 14.36 (0.06) & 11.45 (0.13) & 9.00 (0.33)&  1  &      &  F0  & \\                   
46 & J013442.14+303216.0 & 16.03 (0.08) &15.23 ( $-$ )& 14.56 (U)  &  13.56 (0.04) & 13.02 (0.05) &13.67 (0.03) & 12.93 (0.03) & 10.31 (0.05) & 8.01 (0.14)&  2  & G0Ia &  F8  & GTC \\                   
47 & J013422.40+304835.5 & 15.69 (0.06) &15.29 (0.09) & 15.28 (0.13) &      $-$      &     $-$      &      $-$    &      $-$     &      $-$     &     $-$    &     &      &  A8  & \\                   
48 & J013355.36+304358.2 & 16.48 (0.11) &15.92 (0.17) & 15.70 (U)  &  15.95 (0.10) & 16.20 (0.09) &      $-$    &      $-$     &      $-$     &     $-$    &     &      &      & M01 \\                   
49 & J013303.40+303051.2 & 16.23 (0.08) &16.08 (0.16) & 15.36 (0.15) &  15.04 (0.10) & 14.93 (0.09) &15.17 (0.06) & 14.64 (0.07) & 11.50 (0.15) & 9.57 (U) &  0  &      &  F5  & \\                       
50 & J013408.52+304708.9 & 16.26 (0.09) &16.11 (0.19) & 15.92 (U)  &  16.00 (0.07) & 15.86 (0.06) &      $-$    &      $-$     &      $-$     &     $-$    &     &      &  F8  & \\                   
51 & J013339.46+303126.9 &   $-$          &  $-$      &   $-$        &      $-$      &     $-$      &      $-$    &      $-$     &      $-$     &     $-$    &     &      &      & M01\\                       
52 & J013231.94+303516.7 & 15.82 (0.07) &15.51 (0.10) & 15.33 (0.14) &      $-$      &     $-$      &14.51 (0.03) & 14.14 (0.04) &  9.83 (0.04) & 7.48 (0.11)&  0  &      &  F2  & \\                       
53 & J013439.73+304406.6 & 16.26 (0.09) &16.00 (0.17) & 15.82 (0.20) &      $-$      &     $-$      &      $-$    &      $-$     &      $-$     &     $-$    &     &      &  A8  & \\                       
\end{tabular}                                                                                                                                                                                            
} 
}
\end{sidewaystable*}

\begin{table*}
\caption[]{DIRECT lightcurves of 21 YSGs in M33.}
\label{tab02}  

{\centering
\begin{tabular}{lcccc}
\hline \hline
LGGS & Filter & JD$-$2450000 (d) & Magnitude (mag) & Uncertainty (mag) \\
\hline
J013349.86+303246.1 & B & 332.8548 & 16.990 & 0.009 \\
J013349.86+303246.1 & B & 365.7995 & 17.036 & 0.008 \\
J013349.86+303246.1 & B & 688.9693 & 16.947 & 0.005 \\
\hline
\end{tabular} 
\tablefoot{Table \ref{tab02} is available in its entirety in a machine-readable form at the CDS. A portion, which contains \\ the first three $B-$band measurements of Star 1, is shown here for guidance regarding its form and content.}
} 
\end{table*}

\begin{table*}
\caption[]{Log of spectroscopic observations.}
\label{tab03}  

{\centering
\begin{tabular}{lccccccc}
\hline \hline
ID  & JD & Band & Mag & Exp. time (s) &   SNR   &    Airmass  & Instrument\\ 
\hline
1 & 2456888.59272  &  $V$  & 16.01  &  100  &  70  &  1.2  & OSIRIS/GTC\\ 
2 & 2456888.61312  &  $V$  & 16.19  &  240  &  60  &  1.1  & OSIRIS/GTC\\
3 & 2456888.60277  &  $V$  & 15.99  &  200  & 120  &  1.2  & OSIRIS/GTC\\
44 & 2456888.66294  &  $V$  & 17.28  &  600  & 100  &  1.0  & OSIRIS/GTC\\
46 & 2456888.64362  &  $V$  & 17.34  &  600  &  90  &  1.1  & OSIRIS/GTC\\	
2 & 2457300.91536  &  $K_{s}$  &  14.21  &  3600  &  15  &   1.0  & IRCS/Subaru\\
B324                & 2457301.01534  &  $K_{s}$  &  13.28  &   900  &  75  &   1.1  & IRCS/Subaru\\
\hline
\end{tabular} 
} 
\end{table*}

\begin{table*}
\caption[]{Radial velocities, equivalent widths and generic indices of selected spectral features. The last two columns list measurements of $M_{V}$ and log~$L/L_{\odot}$, based on the calibration of the EW$_{\ion{O}{I}}$. Uncertainties (1$\sigma$) are in parentheses.}
\label{tab04}  
{\centering
\begin{tabular}{crrrrcc|cc}
\hline \hline
ID  & EW$_{H_\alpha}$ ($\AA$) & RV$_{H_\alpha}$ (km~s$^{-1}$)  & EW$_{\ion{O}{I}}$ ($\AA$) & RV$_{\ion{O}{I}}$ (km~s$^{-1}$)  & CaT* & PaT & $M_{V}$ (mag) & log~$L/L_{\odot}$ \\
\hline
1  &   2.8~(0.1)    &  $-$243~(15)  &    2.0~(0.1)  & $-$207~(12) &  15.03~(0.03)  &  2.44~(0.03)  & -8.4 (0.9) & 5.3 (0.4) \\
2  &  $-$9.9~(0.2)  &  $-$320~(5)   &    2.6~(0.2)  & $-$284~(12) &       --       &    --         & -9.6 (1.3) & 5.8 (0.5) \\
3  &   2.9~(0.1)    &  $-$275~(9)   &    2.4~(0.1)  & $-$280~(4)  &   5.18~(0.02)  &  5.56~(0.02)  & -9.3 (1.1) & 5.6 (0.5) \\ 
44  & $-$49.7~(0.1)  &  $-$105~(5)   &    2.1~(0.1)  & $-$203~(8)  &       --       &    --        & -8.6 (1.0) & 5.4 (0.4) \\
46  & $-$46.2~(0.1)  &  $-$170~(5)   &        --     &       --    &  10.82~(0.02)  &  1.69~(0.02) &    --      &       --  \\
\hline
\end{tabular} 
} 
\end{table*}

\begin{table*}
\caption[]{Spectral energy distribution fit parameters of 21 YSGs with available light curves from DIRECT. Uncertainties (1$\sigma$) are in parentheses.}
\label{tab05}  
{\centering
\begin{tabular}{c|cc|c|ccccc}
\hline \hline
           &    \multicolumn{2}{c|}{\cite{Drout12}}    &  $E(B-V)$ & \multicolumn{5}{c}{SED fitting} \\
  ID       & $T_{\text{eff}}$ (K)  & log~$L/L_{\odot}$ &     mag    & $T_{\text{eff}}$ (K) &  $R/R_{\odot}$ & \multicolumn{2}{c}{log~$L/L_{\odot}$} & $T_{dust}$ (K) \\ 
           &                       &                   &            &                      &                &          8 $\mu$m  & 22 $\mu$m        &                \\
\hline
 1 & 5260 & 5.64 &   0.13 (0.01)   & 5500 (70)  &  710 (25) &  5.65            &  5.68 (0.04)  &  670 + 180  \\                   
 2 & 4897 & 5.63 &   0.13 (0.01)   & 5500 (70)  &  685 (25) &  5.59            &  5.61 (0.04)  &  830 + 160  \\    
 4 & 5623 & 5.51 &   0.06 (0.01)   & 5750 (70)  &  510 (15) &  \multicolumn{2}{c}{5.41 (0.04)} &  870 + 110  \\    
 6 & 5701 & 5.40 &   0.19 (0.01)   & 6000 (80)  &  485 (20) &  5.44            &  5.45 (0.04)  & 1270 + 130  \\    
13 & 6295 & 5.30 &   0.12 (0.02)   & 6540 (110) &  360 (15) &  5.35            &  5.48 (0.04)  &  600 + 150  \\    
14 & 5333 & 5.28 &   0.14 (0.01)   & 5800 (120) &  450 (25) &  \multicolumn{2}{c}{5.31 (0.04)} &    --       \\    
15 & 5597 & 5.27 &   0.13 (0.01)   & 5960 (110) &  400 (20) &  5.26            &  5.28 (0.04)  &    110      \\            
17 & 5997 & 5.26 &   0.13 (0.01)   & 6250 (70)  &  370 (10) &  \multicolumn{2}{c}{5.27 (0.04)} &    --       \\    
22 & 6807 & 5.23 &   0.03 (0.02)   & 6290 (110) &  320 (15) &  \multicolumn{2}{c}{5.16 (0.04)} &    --       \\    
26 & 5116 & 5.21 &   0.13 (0.01)   & 5440 (130) &  400 (25) &  \multicolumn{2}{c}{5.10 (0.04)} &    --       \\    
27 & 8472 & 5.20 &   0.13 (0.01)   & 7920 (140) &  220 (10) &  \multicolumn{2}{c}{5.23 (0.03)} &    --       \\    
29 & 5296 & 5.19 &   0.10 (0.02)   & 5670 (140) &  385 (25) &  \multicolumn{2}{c}{5.16 (0.04)} &   1040      \\    
31 & 6067 & 5.17 &   0.13 (0.01)   & 6250 (70)  &  340 (10) &  \multicolumn{2}{c}{5.21 (0.04)} &   1000      \\    
32 & 6053 & 5.16 &   0.13 (0.01)   & 6250 (70)  &  340 (10) &  5.21            &  5.22 (0.04)  & 1200 + 160  \\    
33 & 6807 & 5.16 &   0.11 (0.02)   & 6730 (100) &  285 (10) &  5.24            &  5.27 (0.04)  & 170 + winds \\    
36 & 5584 & 5.15 &   0.07 (0.01)   & 5750 (70)  &  345 (10) &  \multicolumn{2}{c}{5.07 (0.04)} &    --       \\    
38 & 6011 & 5.14 &   0.06 (0.01)   & 6000 (70)  &  325 (10) &  \multicolumn{2}{c}{5.08 (0.04)} &    --       \\    
39 & 8394 & 5.12 &   0.13 (0.01)   & 8240 (100) &  190  (5) &  \multicolumn{2}{c}{5.16 (0.03)} &    --       \\    
40 & 7998 & 5.12 &   0.13 (0.01)   & 7700 (410) &  215 (15) &  \multicolumn{2}{c}{5.18 (0.04)} &   winds?    \\    
48 & 7294 & 5.07 &   0.14 (0.07)   & 7230 (350) &  235 (15) &  \multicolumn{2}{c}{5.15 (0.07)} &   1230?     \\    
51 & 7998 & 5.06 &   0.12 (0.01)   & 7750 (70)  &  200  (5) &  \multicolumn{2}{c}{5.11 (0.03)} &    --       \\    
\hline
\end{tabular}                                                                                   
} 
\end{table*}

\begin{table*}
\caption[]{Spectral energy distribution fit parameters of three dusty YSGs with available spectroscopy from GTC. \\ Uncertainties (1$\sigma$) are in parentheses. The asterisk indicates the value of reddening adopted for star 44 from \cite{Humph13}.}
\label{tab06}  
{\centering
\begin{tabular}{c|cc|c|ccccc}
\hline \hline
           &    \multicolumn{2}{c|}{\cite{Drout12}}    &  $E(B-V)$   & \multicolumn{5}{c}{SED fitting} \\
  ID       & $T_{\text{eff}}$ (K)  & log~$L/L_{\odot}$ &     mag     & $T_{\text{eff}}$ (K) &  $R/R_{\odot}$ & \multicolumn{2}{c}{log~$L/L_{\odot}$} & $T_{dust}$ (K) \\ 
           &                       &                   &             &                      &                &          8 $\mu$m  & 22 $\mu$m        &                \\
\hline             
3  & 6980 & 5.53 &  0.11 (0.01)         & 7500 (70)    & 355 (10) &   5.59 &  5.62 (0.03) &     1200 + 140      \\                
44 & 5450 & 5.10 &  0.16 (0.02) / 0.50*  & 6400 (240)   & 490 (20) &   5.65 &  5.72 (0.06) &  980 + 380 + 130    \\                
46 & 5430 & 5.08 &  0.13 (0.01)         & 5610 (140)   & 375 (25) &   5.20 &  5.25 (0.04) &  950 + 320 + 100    \\             
\hline
\end{tabular} 
} 
\end{table*}

\begin{table*}
\caption[]{Binned $BVI_{c}$ photometry and inferred parameters of star 2. Uncertainties (1$\sigma$) are in parentheses.}
\label{tab07}  
{\centering
\begin{tabular}{c|ccc|ccc}
\hline \hline
  JD     &    $B$       &       $V$    &     $I_{c}$  &  $A_{V}$ fixed & \multicolumn{2}{c} {$A_{V}$ free} \\
         &  (mag)       &     (mag)    &     (mag)    & $T_{\text{eff}}$ (K) &  $T_{\text{eff}}$ (K) &  $A_{V}$ (mag)\\ 
\hline
333.6281 & 16.836 (0.020) & 16.083 (0.038) & 15.277 (0.019) &  5750 (70)  &  5670 (170) &  0.29 (0.20) \\                                      
365.3147 & 16.776 (0.014) & 16.041 (0.031) & 15.260 (0.033) &  5750 (70)  &  5660 (150) &  0.22 (0.17) \\                                     
686.6760 & 17.026 (0.020) & 16.115 (0.047) & 15.237 (0.005) &  5500 (70)  &  5400 (150) &  0.21 (0.18) \\                                     
689.9840 & 16.967 (0.014) & 16.115 (0.029) & 15.257 (0.026) &  5570 (130) &  5520 (100) &  0.28 (0.10) \\                                     
695.4098 & 17.009 (0.022) & 16.147 (0.014) & 15.242 (0.004) &  5500 (70)  &  5540 (120) &  0.40 (0.12) \\                                     
730.8744 & 16.949 (0.012) & 16.094 (0.027) & 15.236 (0.009) &  5550 (120) &  5510 ( 85) &  0.27 (0.06) \\                                     
1379.632 & 16.744 (0.011) & 15.931 (0.051) & 15.008 (0.021) &  5500 (70)  &  5940 (550) &  0.80 (0.46) \\                                     
1429.511 & 16.721 (0.008) & 15.864 (0.060) & 14.987 (0.007) &  5500 (70)  &  5590 (150) &  0.41 (0.18) \\                                     
1433.367 & 16.733 (0.002) & 15.846 (0.053) & 14.991 (0.004) &  5500 (70)  &  5450 (150) &  0.24 (0.17) \\                                     
1459.135 & 16.734 (0.037) & 15.937 (0.038) & 14.964 (0.060) &  5510 (80)  &  6310 (640) &  1.21 (0.53) \\                                     
1462.911 & 16.751 (0.030) & 15.921 (0.031) & 14.970 (0.039) &  5500 (70)  &  5850 (380) &  0.79 (0.33) \\                                     
1483.052 & 16.883 (0.039) & 15.991 (0.053) & 15.014 (0.065) &  5490 (90)  &  5700 (350) &  0.72 (0.36) \\                                     
1485.657 & 16.921 (0.015) & 15.979 (0.029) & 15.026 (0.013) &  5500 (70)  &  5470 (110) &  0.47 (0.12) \\                                     
2517.839 & 17.193 (0.002) & 15.961 (0.029) & 14.889 (0.040) &  5010 (90)  &  5050 (120) &  0.39 (0.15) \\                                     
2531.756 & 17.268 (0.001) & 15.999 (0.012) & 14.945 (0.041) &  5000 (80)  &  5000 ( 70) &  0.32 (0.05) \\                                     
2557.851 & 17.336 (0.008) & 16.052 (0.008) & 14.789 (0.152) &  4720 (200) &  5260 (190) &  1.03 (0.36) \\                                     
2589.811 & 17.360 (0.001) & 16.047 (0.008) & 14.948 (0.007) &  5000 (70)  &  5000 ( 70) &  0.43 (0.01) \\                                     
2646.644 & 17.339 (0.004) & 16.015 (0.003) & 14.952 (0.013) &  5000 (70)  &  4820 (130) &  0.17 (0.13) \\                                     
2662.621 & 17.415 (0.003) & 16.060 (0.016) & 15.011 (0.007) &  5000 (70)  &  4750 ( 70) &  0.09 (0.02) \\                                     
2908.821 & 17.261 (0.003) & 16.048 (0.007) & 14.973 (0.022) &  5000 (80)  &  5040 (110) &  0.37 (0.12) \\                                     
2932.817 & 17.103 (0.002) & 16.004 (0.002) & 15.024 (0.039) &  5250 (70)  &  5240 ( 90) &  0.34 (0.09) \\                                     
2958.827 & 17.201 (0.013) & 15.986 (0.012) & 14.987 (0.049) &  5220 (110) &  5000 ( 70) &  0.17 (0.06) \\                                     
3000.654 & 17.173 (0.013) & 15.970 (0.016) & 14.967 (0.039) &  5240 (90)  &  5000 ( 70) &  0.17 (0.05) \\                                     
3567.580 & 16.752 (0.028) & 15.776 (0.149) & 14.922 (0.121) &  5510 (100) &  5470 (410) &  0.33 (0.30) \\                                     
\hline
\end{tabular} 
} 
\end{table*}


\begin{appendix}

\onlfig{
\begin{figure*} 
\includegraphics[width=7in]{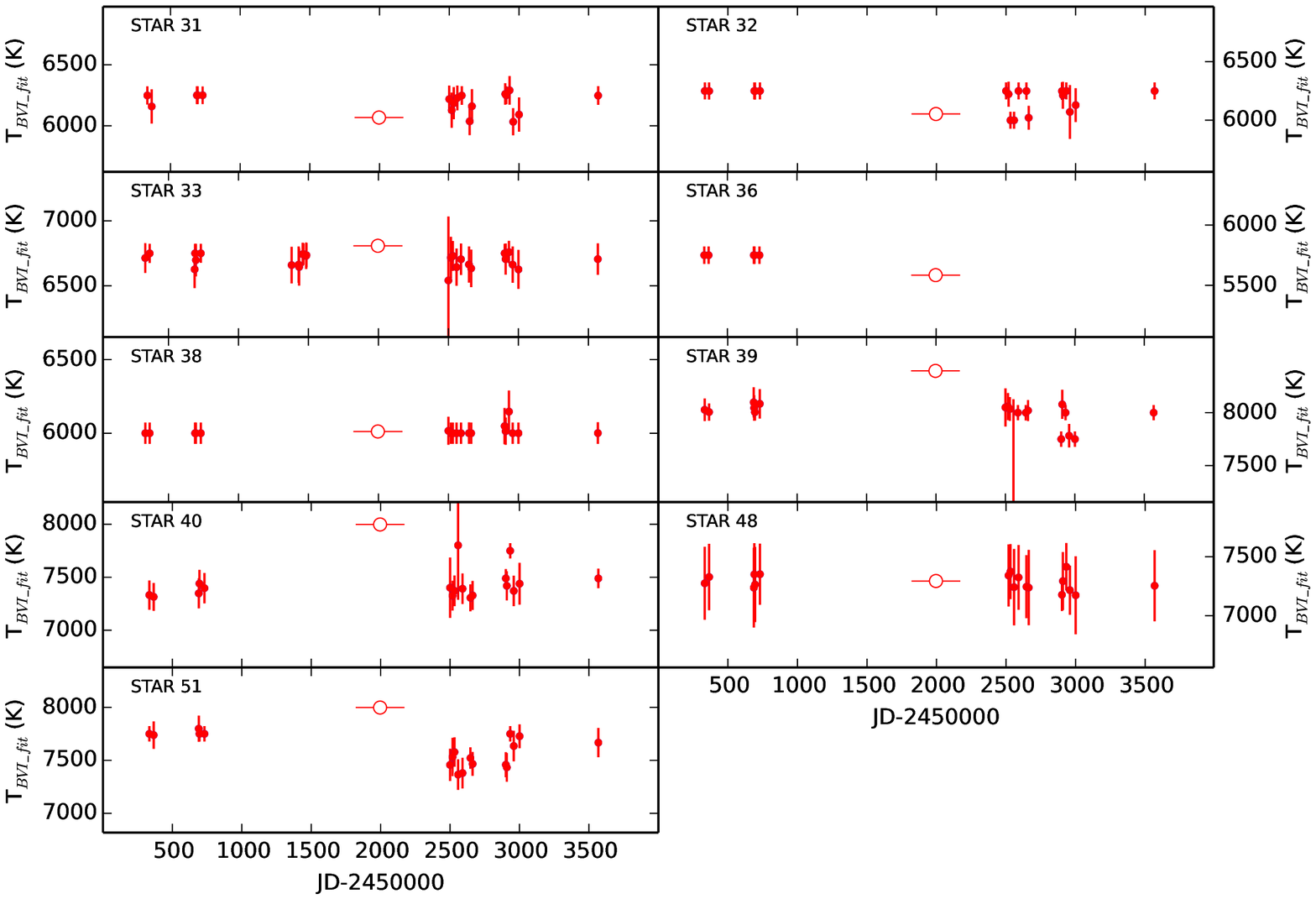}
\caption{Same as in Fig. \ref{fig07}, but for the last 9 stars from Table \ref{tab05}.}
\end{figure*}

\begin{figure*} 
\includegraphics[width=7.5in]{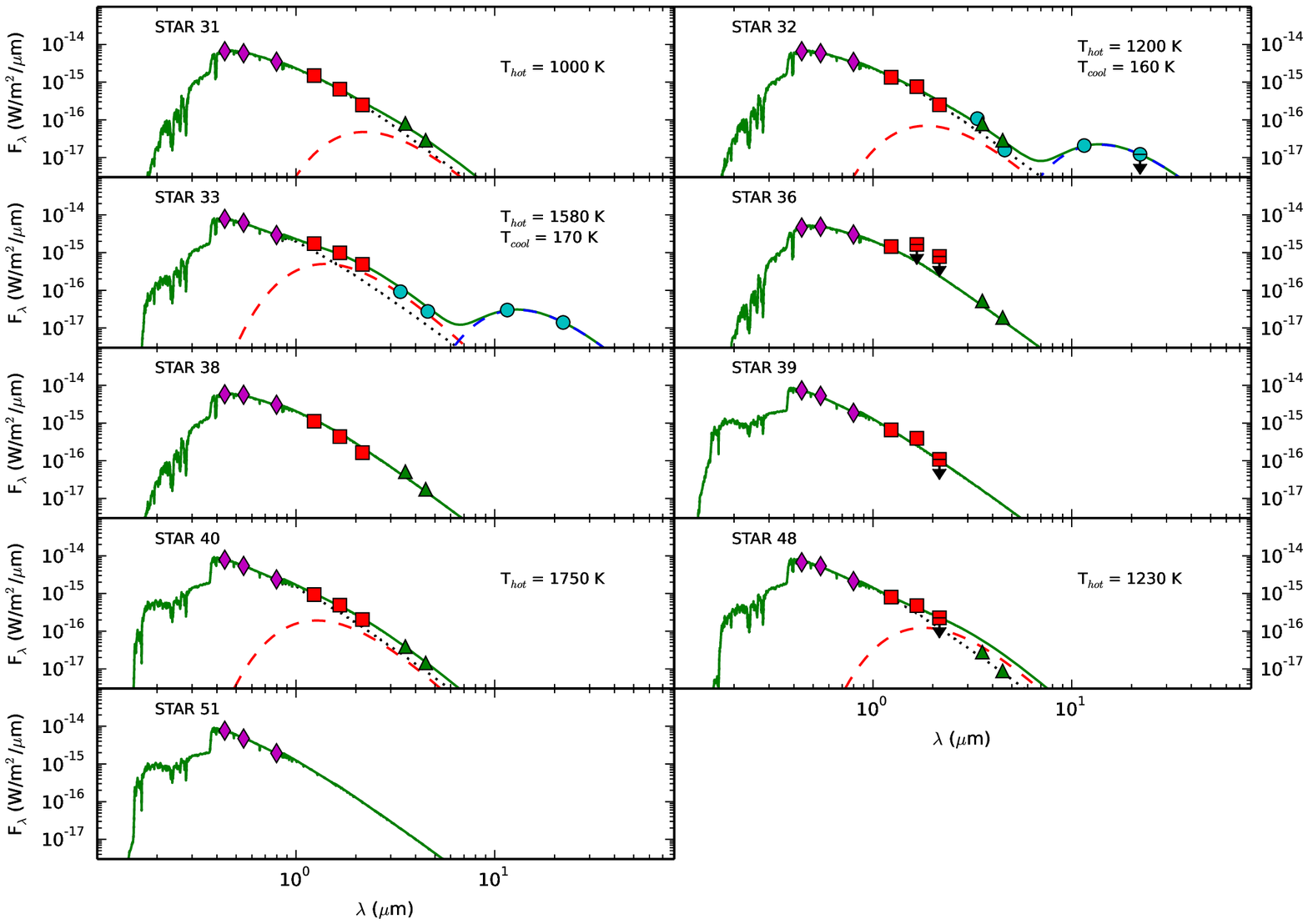}
\caption{Same as in Fig. \ref{fig08}, but for the last 9 stars from Table \ref{tab05}.}
\end{figure*}
}

\end{appendix}

\end{document}